\documentclass[final]{IEEEtran}
\usepackage{amsmath,amssymb,graphicx,epstopdf,enumitem,algorithmic,pgfplots,subfig,cite}
\usetikzlibrary{shapes,arrows,positioning}

\long\def\comment#1{}

\newfont{\bbb}{msbm10 scaled 700}

\newfont{\bb}{msbm10 scaled 1100}

\newcommand{\av}{{\bf a}}

\newcommand{\nv}{{\bf n}}

\newcommand{\rv}{{\bf r}}
\newcommand{\sv}{{\bf s}}
\newcommand{\tv}{{\bf t}}
\newcommand{\uv}{{\bf u}}
\newcommand{\wv}{{\bf w}}
\newcommand{\vv}{{\bf v}}
\newcommand{\xv}{{\bf x}}
\newcommand{\yv}{{\bf y}}
\newcommand{\zv}{{\bf z}}

\newcommand{\Am}{{\bf A}}
\newcommand{\Bm}{{\bf B}}

\newcommand{\Gm}{{\bf G}}

\newcommand{\Id}{{\bf I}}
\newcommand{\Jm}{{\bf J}}

\newcommand{\Nm}{{\bf N}}

\newcommand{\Rm}{{\bf R}}
\newcommand{\Sm}{{\bf S}}
\newcommand{\Tm}{{\bf T}}

\newcommand{\Wm}{{\bf W}}
\newcommand{\Vm}{{\bf V}}
\newcommand{\Xm}{{\bf X}}
\newcommand{\Ym}{{\bf Y}}
\newcommand{\Zm}{{\bf Z}}

\newcommand{\Fc}{{\cal F}}
\newcommand{\Gc}{{\cal G}}
\newcommand{\Hc}{{\cal H}}

\newcommand{\Jc}{{\cal J}}

\newcommand{\Lc}{{\cal L}}

\newcommand{\Nc}{{\cal N}}

\newcommand{\Gammam}{\hbox{\boldmath$\Gamma$}}

\newcommand{\diag}{{\hbox{diag}}}

\newcommand{\herm}{{\sf H}}
\newcommand{\trasp}{{\sf T}}

\newcommand{\transp}{{\sf T}}

\newcommand{\Gfd}{\mbox{$\boldsymbol{\mathcal{G}}$}}
\newcommand{\Hfd}{\mbox{$\boldsymbol{\mathcal{H}}$}}

\newcommand{\Nfd}{\mbox{$\boldsymbol{\mathcal{N}}$}}

\DeclareMathOperator{\arctanh}{arctanh}

\newsavebox{\ieeealgbox}
\newenvironment{boxedalgorithmic}
  {\begin{lrbox}{\ieeealgbox}
   \begin{minipage}{\dimexpr\columnwidth-2\fboxsep-2\fboxrule}
   \begin{algorithmic}}
  {\end{algorithmic}
   \end{minipage}
   \end{lrbox}\noindent\fbox{\usebox{\ieeealgbox}}}

\newlength\figureheight
\newlength\figurewidth

\begin{document}

\title{Blind Source Separation Algorithms Using Hyperbolic and Givens Rotations for High-Order QAM Constellations}

\author{Syed~A.~W.~Shah,~\IEEEmembership{Student~Member,~IEEE,}
        Karim~Abed-Meraim$^\ast$,~\IEEEmembership{Senior~Member,~IEEE,}
        and~Tareq~Y.~Al-Naffouri,~\IEEEmembership{Member,~IEEE,}%
\thanks{S. A. W. Shah and T. Y. Al-Naffouri are with the EE department, King Fahd University of Petroleum and Minerals, Dhahran, Saudi Arabia (e-mail: awaiswahab100@gmail.com). K. Abed-Meraim is with the University of Orl\'{e}ans, PRISME Lab. Orl\'{e}ans, France (e-mail: karim.abed-meraim@univ-orleans.fr). T. Y. Al-Naffouri is also associated with the EE department, King Abdullah University of Science and Technology, Thuwal, Saudi Arabia (e-mail: tareq.alnaffouri@kaust.edu.sa).}}

\markboth{Submitted to IEEE Trans. Signal Process., June~2016}
{Shell \MakeLowercase{\textit{Syed et al.}}: Bare Demo of IEEEtran.cls for Journals}

\maketitle

\begin{abstract}
This paper addresses the problem of blind demixing of instantaneous mixtures in a multiple-input multiple-output communication system. The main objective is to present efficient blind source separation (BSS) algorithms dedicated to moderate or high-order QAM constellations. Four new iterative batch BSS algorithms are presented dealing with the multimodulus (MM) and alphabet matched (AM) criteria. For the optimization of these cost functions, iterative methods of Givens and hyperbolic rotations are used. A pre-whitening operation is also utilized to reduce the complexity of design problem. It is noticed that the designed algorithms using Givens rotations gives satisfactory performance only for large number of samples. However, for small number of samples, the algorithms designed by combining both Givens and hyperbolic rotations compensate for the ill-whitening that occurs in this case and thus improves the performance. Two algorithms dealing with the MM criterion are presented for moderate order QAM signals such as 16-QAM. The other two dealing with the AM criterion are presented for high-order QAM signals. These methods are finally compared with the state of art batch BSS algorithms in terms of signal-to-interference and noise ratio, symbol error rate and convergence rate. Simulation results show that the proposed methods outperform the contemporary batch BSS algorithms.
\end{abstract}


\IEEEpeerreviewmaketitle

\section{Introduction}\label{sec:intro}
\IEEEPARstart{B}{lind} source separation (BSS) is a fundamental signal processing technology that has been intensively used in many systems including biomedical, audio and industrial applications \cite{handbook}. In the context of overdetermined multiple-input multiple-output (MIMO) systems, BSS aims to find a separation matrix using the received signals and \emph{a priori} information about the statistics or the nature of transmitted source signals. Usually, in a communication system, the modulation technique being used is known \emph{a priori}. One can utilize such information to recover the same properties in the output signal and thus estimate the source signal blindly. Various BSS cost functions can be found in literature \cite{handbook, haykin} depending upon the types of source signals. Among them, the constant modulus (CM) criterion for phase/frequency modulated signals such as PSK/FSK and multimodulus (MM) criterion for QAM signals have attracted great interest.

The CM criterion \cite{CMAcost} restricts the squared modulus of the output to be a constant, but such algorithms even work for non CM signals. They lead to a number of constant modulus algorithms (CMA) used for blind equalization \cite{johnsonCMAeq, AbrarCMA}, blind beamforming \cite{GoochBeamforming, veenbook} and BSS \cite{acma, GandHGCMA}. On the other hand, the MM criterion \cite{MMcost} takes into account the knowledge of square QAM constellation. Its respective cost function deals with the real and imaginary parts of the signal separately and leads to numerous multimodulus algorithms (MMA) used for the application of blind equalization \cite{MMcost} and BSS \cite{MIMOMMA, AMMA, AwaisICASSP15}. MMA outperforms the CMA for the case of square QAM, which is used in many modern communication systems such as LTE \cite{LTEref} and WiMAX \cite{WiMaxref}. For such advanced systems requiring high data rate, high-order modulations having better spectral efficiency are used such as 64-QAM. For these high-order modulations, MMA leads to considerable amount of residual errors and does not ensure low symbol error rate (SER). Thus, in order to improve the performance of BSS algorithms for high-order QAM signals, a number of alphabet matched (AM) penalty terms \cite{Li1997, Barbarossa1997, HeAMA2004} were suggested. All of these AM cost functions were found to have good local convergence properties and therefore require a good initialization \cite{HeAMA2004}. Thus, alphabet matched algorithms (AMA) should be used along with either CMA or MMA, as both of them have good global convergence properties.

Out of numerous CMA solutions, the algebraic one named Analytical Constant Modulus Algorithm (ACMA) \cite{acma} provides an exact separation in the noise-free case. To overcome the drawback of numerical complexity of ACMA, two batch BSS algorithms namely Givens CMA (G-CMA) and Hyperbolic G-CMA (HG-CMA) were presented in \cite{GandHGCMA}, which outperform ACMA. The adaptive versions of ACMA and G-CMA were presented in \cite{adaptiveACMA} and \cite{adaptiveGCMA}, respectively. Similarly, for the MM criterion, an adaptive MMA algorithm was presented in \cite{MIMOMMA}, which outperforms the Multi-User Kurtosis (MUK) algorithm \cite{MUK}. Seeing the popularity of ACMA, the same analytical approach was used for MM signals and thus an Analytical Multimodulus Algorithm (AMMA) was presented in \cite{AMMA}. In terms of blind equalization considering a single source, a number of AMA were presented using a combination of CM/MM and AM cost functions either in hybrid \cite{HeAMA2004} or dual mode \cite{AmaDualMode}. It is shown in \cite{Beasley2006} that hybrid and dual mode have nearly the same performance. An adaptive blind equalization algorithm by combining CM and AM cost functions was presented in \cite{Moazzami2008} which separates all the sources using multi-stage cascaded equalizers, where the number of equalizers were equal to the number of sources.

\subsection{Contributions}\label{Contribution}
In this paper, we propose four new batch BSS algorithms utilizing the MM and AM criteria for MIMO systems. The major contribution includes the optimization of MM/AM criterion using Givens and hyperbolic rotation parameters for the case of multiple sources. Two algorithms are designed by minimization of MM and AM cost functions using real Givens rotations and named as Givens MMA (G-MMA) and Givens AMA (G-AMA), respectively. The other two algorithms are designed using both real Givens and hyperbolic rotations and thus named as Hyperbolic G-MMA (HG-MMA) and Hyperbolic G-AMA (HG-AMA). To the best of our knowledge, this is the first paper which presents batch BSS algorithms for MM and AM criteria.

Previously, stochastic gradient techniques were used for the minimization of MM and AM criteria \cite{MIMOMMA, Li1997, Barbarossa1997, HeAMA2004, Moazzami2008}, thus all of these algorithms are adaptive and slow in convergence. Therefore, we compare our algorithms with batch BSS algorithms designed for CM signals such as ACMA, G-CMA and HG-CMA, in terms of signal to interference and noise ratio (SINR), symbol error rate (SER), and convergence rate.

\subsection{Paper Organization and Notations}\label{org_not}
This paper is organized as follows. In Section \ref{sec:Prob_Form}, the data model and a brief overview of BSS principle are presented. Section \ref{sec:Algo_Design} defines the used criteria as well as Givens and hyperbolic rotations. The derivation of proposed algorithms G-MMA, HG-MMA, G-AMA and HG-AMA is presented in Sections \ref{sec:GMMA}, \ref{sec:HGMMA}, \ref{sec:GAMA_Algo}, and \ref{sec:HGAMA_Algo}, respectively. Section \ref{sec:comments} includes some comments to highlight the important features of the proposed algorithms. Simulation results are presented in Section \ref{sec:Sim_Res} and Section \ref{sec:conclusion} concludes the paper.

Following are the notations used in this paper. $\xv$ denotes a column vector where its $i$th entry is denoted by $x_i$. The real and imaginary parts of $x$ are denoted by $x_R$ and $x_I$. The matrix and its ($i,j$)-th entry are denoted by $\Xm$ and $x_{ij}$, respectively. If the matrix consists of only real elements then it is represented as $\acute{\Xm}$. $\Id$ represents the identity matrix. $(.)^\transp$ and $(.)^\herm$ are used to represent matrix/vector transpose and complex conjugate transpose, respectively. $\underline{x}$ denotes the pre-filtered variables. $\iota$ is used to denote $\sqrt{-1}$. $E[.]$ is the mathematical expectation operator and $|.|$ denotes the modulus function.

\section{Problem Formulation}\label{sec:Prob_Form}
Consider a MIMO system consisting of $N_t$ sources, each having a single antenna element and a receiver equipped with an array of $N_r$ antennas. All sources transmit their signals over the same band of frequencies. Each transmitted source signal $s(i) = s_{R}(i) + \iota s_{I}(i)$ is drawn from an $L$-ary square QAM constellation where $s_{R}(i), s_{I}(i) \in \left\{ \pm 1, \pm 3, \ldots, \pm (\sqrt{L}-1)  \right\}$. The unknown source signal $\sv(i) = \begin{bmatrix} s_1(i) & \cdots & s_{N_t}(i) \end{bmatrix}^\transp$ is passed through a flat fading channel represented by an unknown mixing matrix $\Am \in \mathbb{C}^{N_r \times N_t}$ whose elements $a_{mn}$ denotes the channel path between transmitter $n$ and receiver $m$. The received signal with the added noise can be mathematically represented as
\begin{equation}\label{Rx_Sig}
    \yv(i) = \Am \sv(i) + \nv(i)
\end{equation}
where $\nv(i) = \begin{bmatrix} n_1(i) & \cdots & n_{N_r}(i) \end{bmatrix}^\transp$ is the white noise vector of covariance $\sigma_n^2 \Id_{N_r}$. Here, we assume that the mixing matrix $\Am$ is of full column rank which implies that $N_r \geq N_t$.

The objective is to recover the source signals $\sv(i)$ without prior knowledge of the channel or without the use of training sequences (pilots). This is accomplished using BSS which relies on the observation vector $\yv(i)$ only and also uses some source's structural information. In order to recover the source signals (up to a permutation and scaling factors \cite{GandHGCMA}), we apply a $(N_t \times N_r)$ separation matrix $\Wm$ according to
\begin{equation}\label{SMRecOut}
    \zv(i) = \Wm \yv(i) = \Wm \Am \sv(i)+\Wm \nv(i) = \Gm \sv(i)+\bar{\nv}(i)
\end{equation}
where $\zv(i)=\begin{bmatrix} z_1(i) & \cdots & z_{N_t}(i) \end{bmatrix}^\trasp$ is the estimated source signal vector, $\Gm = \Wm \Am$ is the $(N_t \times N_t)$ global system matrix and $\bar{\nv}(i) = \Wm \nv(i)$ is the filtered noise vector.

In this paper, we consider batch BSS algorithms in which $N_s$ samples of the received signal are collected and then a separation matrix is applied on the received data packet $\Ym = \begin{bmatrix} \yv(1) & \cdots & \yv(N_s) \end{bmatrix}$, so that \eqref{Rx_Sig} and \eqref{SMRecOut} can be rewritten as
\begin{equation}\label{SMRecInK}
    \Ym = \Am \Sm + \Nm, \quad \Zm = \Wm \Ym
\end{equation}
where $\Zm$, $\Sm$ and $\Nm$ are defined in a way similar to the definition of $\Ym$. In what follows, we seek a separation matrix in the form $\Wm=\Vm \Bm$, where $\Bm$ is a $(N_t \times N_r)$ pre-whitening matrix that can be computed from a covariance matrix as in \cite{veenbook}, or simply a $(N_t \times N_r)$ projection matrix onto the signal subspace (since pre-whitening is needed only for the G-MMA/G-AMA but not for the HG-MMA/HG-AMA methods). Our main contribution lies in designing efficient methods for the computation of the matrix $\Vm$ in order to minimize cost functions suitable for high-order QAM signals.

\section{Algorithm Design}\label{sec:Algo_Design}
The first step of algorithm design is the selection of a suitable cost function. Various cost functions can be found in the literature depending upon the properties/types of source signals. In the considered case of square QAM signals, we have selected the following cost functions.

\subsection{Cost Functions}\label{sec:Cost_Func}
For low order square signals we design MMA using MM cost function, however for high-order square QAM signals, we design AMA using AM cost function as shown next.

\subsubsection{Multimodulus (MM) Cost Function}
For multimodulus signals e.g., square QAM, one proposes to estimate the matrix $\Vm$ by minimizing the MM criterion defined in \cite{MMcost} as
\begin{equation}\label{MMcost}
    \Jc_{\text{MMA}}(\Vm) \!=\! \sum_{j=1}^{N_t} \mathbb{E} \left[ \left(z_{j,R}^2(i)-R_R\right)^2 + \left(z_{j,I}^2(i)-R_I\right)^2 \right]
\end{equation}
where $R_R=R_I=\mathbb{E}[|s_R(i)|^4]/\mathbb{E}[|s_R(i)|^2]$ are dispersion constants of the real and the imaginary parts, respectively. This cost function was designed such that its minimization can be interpreted as fitting the signal into a square shaped signal. Thus, it contains structural information of QAM signals and also has an inherent ability to restore the phase of the signal. Moreover, the MM cost function has several advantages over the CM one \cite{MMAequa} and leads to: i) faster convergence algorithms \cite{PicchiPrati87,yang2002multimodulus}, ii) carrier phase recovery \cite{YuanTsai05PhaseRecov}, iii) less undesirable minima \cite{li2009performance} and iv) ease in hardware implementation \cite{mizuno2003hardware}.

\subsubsection{Alphabet Matched (AM) Cost Function}\label{AM_Cost_Func}
Out of the variety of AM cost functions \cite{Li1997, Barbarossa1997, Li1995, He2001}, we have selected the one presented in \cite{Amin2001} as
\begin{equation}\label{AMA_Cost}
  \Jc_{\text{AMA}}(\Vm) = \sum_{j=1}^{N_t} \mathbb{E} \left[ g(z_{j,R}(i)) + g(z_{j,I}(i)) \right]
\end{equation}
where $g(x)$ is the constellation matched error (CME) term defined as
\begin{equation}\label{CMETerm}
    g(x) = 1 - \sin^{2n}(x\frac{\pi}{2d})
\end{equation}
where $n \in \mathbb{N}$ and $2d$ is the minimum distance between alphabet points. The CME in \eqref{CMETerm} satisfies a number of properties that shape the high-order square QAM signals including: i) it does not favor alphabet members over others, thus it has a uniform behavior, ii) it is locally symmetric around each alphabet point, and iii) it places the highest penalty at the maximum deviation i.e., the midpoint between two alphabet points and does not place any penalty for zero errors i.e., at the alphabet points.

The next step is to devise an efficient method for the optimization of the previous cost functions. To guarantee a fast convergence with relatively easy implementation, we propose to decompose the separation matrix $\Vm$ into a product of elementary rotations, similar to Jacobi-like algorithms \cite{JacobiSIAM, matrixcomp}, used for matrix diagonalization. Hence $\Vm$ is derived using a sequence of Givens and hyperbolic rotations, whose parameters are computed by minimizing the MM/AM criteria.

\subsection{Review of Givens and Hyperbolic (Shear) Rotations}
\subsubsection{Givens Rotations}
The unitary Givens rotation $\Gfd_{p,q}(\theta, \alpha)$ is an $(m \times m)$ identity matrix except for the four entries $\Gc_{pp}, \Gc_{qq}, \Gc_{pq}$ and $\Gc_{qp}$ given by
\begin{equation}\label{GivensMatrix}
    \begin{bmatrix}
        \Gc_{pp} & \Gc_{pq}\\
        \Gc_{qp} & \Gc_{qq}\\
    \end{bmatrix}
    =
    \begin{bmatrix}
        \cos(\theta) & \mathrm{e}^{\iota \alpha} \sin(\theta)\\
        -\mathrm{e}^{-\iota \alpha} \sin(\theta) & \cos(\theta)\\
    \end{bmatrix}
\end{equation}
where $\theta \in \left[ -\pi/2, \pi/2 \right]$ and $\alpha \in \left[ -\pi/2, \pi/2 \right]$ are angle parameters with $\alpha = 0$ for the real case.

\subsubsection{Hyperbolic Rotations}
The non-unitary Hyperbolic rotation $\Hfd_{p,q}(\gamma, \beta)$ is an $(m \times m)$ identity matrix, except for the four elements $\Hc_{pp}, \Hc_{qq}, \Hc_{pq}$ and $\Hc_{qp}$ given by
\begin{equation}\label{HyperMatrix}
    \begin{bmatrix}
        \Hc_{pp} & \Hc_{pq}\\
        \Hc_{qp} & \Hc_{qq}\\
    \end{bmatrix}
    =
    \begin{bmatrix}
        \cosh(\gamma) & \mathrm{e}^{\iota \beta} \sinh(\gamma)\\
        \mathrm{e}^{-\iota \beta} \sinh(\gamma) & \cosh(\gamma)\\
    \end{bmatrix}
\end{equation}
where $\gamma \in \left[ -\Gamma,\Gamma \right], \Gamma>0$ and $\beta \in \left[ -\pi/2, \pi/2 \right]$. Similar to Givens rotations, in the real case, $\beta = 0$.

\subsection{Motivation for using Real Givens and Hyperbolic Rotations}\label{sec:motivation}
For large number of sources $N_t$, the difficulty to estimate $\Vm$ increases. Thus, to simplify the estimation process, similar to Jacobi-like algorithms \cite{JacobiSIAM, matrixcomp}, we propose to decompose $\Vm$ into a product of $N_t(N_t-1)$ elementary Givens rotations as
\begin{equation}\label{GivensVcomplex}
    \Vm = \prod_{N_{Sweeps}} \prod_{1 \leq p,q \leq N_t} \Gfd_{p,q}(\theta, \alpha)
\end{equation}
where $N_{Sweeps}$ denotes the number of iterations. Parameters $\theta$ and $\alpha$ are computed  in order to minimize the MM criterion (\ref{MMcost}). Consider a unitary transformation $\Zm=\Gfd_{p,q} \underline{\Ym}$, which according to (\ref{GivensMatrix}) only changes the rows `$j=p$' and `$j=q$' of $\underline{\Ym}$ so that
\begin{equation}\label{GivensComplexOnY}
    \begin{gathered}
        \! z_{ji} \!=\! \underline{y}_{ji} ~ \text{for} ~ j \neq p,q, \quad z_{pi} \!=\! \cos(\theta) \underline{y}_{pi} + \mathrm{e}^{\iota \alpha} \sin(\theta) \underline{y}_{qi}\\
        z_{qi}=-\mathrm{e}^{-\iota \alpha} \sin(\theta) \underline{y}_{pi} + \cos(\theta) \underline{y}_{qi}
    \end{gathered}
\end{equation}
By omitting the constant terms of $\Zm$ independent of $(\theta, \alpha)$, (\ref{MMcost}) can be re-written as:
\begin{multline}\label{MMcostGivensComplex}
    \Jc_{\text{MMA}}(\Gfd_{pq})=\sum_{i=1}^{N_s} \left[ \left( z_{pi,R}^2-R_R \right)^2 + \left( z_{qi,R}^2-R_R \right)^2 \right. \\ \left. + \left( z_{pi,I}^2-R_I \right)^2 + \left( z_{qi,I}^2-R_I \right)^2 \right]
\end{multline}
where each term $z_{pi,R}^2$, $z_{qi,R}^2$, $z_{pi,I}^2$, $z_{qi,I}^2$ equals to $g_i^1 \cos(2\theta) + g_i^2 \cos(2\theta) \cos(2\alpha) + g_i^3 \cos(2\theta) \sin(2\alpha) + g_i^4 \sin(2\theta) \cos(\alpha) + g_i^5 \sin(2\theta) \sin(\alpha) + g_i^6 \cos(2\alpha) + g_i^7 \sin(2\alpha) + g_i^8$ and $g_i^j, j=1,\cdots,8$ are constant terms depending upon the entries of $\underline{\Ym}$. As we can see, further analytical simplification and thus the solution of (\ref{MMcostGivensComplex}) is quite complicated. Similar is the case with hyperbolic rotations. These difficulties motivated us to come up with a different solution explained below\footnote{We have presented this work partly in \cite{AwaisICASSP15}.}.

\section{Givens MMA (G-MMA)}\label{sec:GMMA}
Until now, we have been working in the complex domain and to deal with the previously mentioned challenges, we will now work in the real domain. Hence, matrix $\underline{\Ym}$ is converted into a real matrix $\acute{\underline{\Ym}}$ containing real and imaginary parts in separate rows as defined in (\ref{Vstruct}). Moreover, a special structure of matrix $\Vm$ is introduced and maintained while applying the rotations. The transformed real received signal and output signal can now be written as $\acute{\underline{\Ym}}$ and $\acute{\Zm} = \acute{\Vm} \acute{\underline{\Ym}}$, respectively, where
\begin{equation}\label{Vstruct}
    \acute{\underline{\Ym}} \!=\!
    \begin{bmatrix}
        \underline{\Ym}_R\\
        \underline{\Ym}_I\\
    \end{bmatrix}
    (2N_t \times N_s), ~
    \acute{\Vm} \!=\!
    \begin{bmatrix}
        \Vm_R & -\Vm_I\\
        \Vm_I & \Vm_R\\
    \end{bmatrix}
    (2N_t \times 2N_t)
\end{equation}
Similarly, $\acute{\Sm}$ and $\acute{\Zm}$ are now $(2N_t \times N_s)$ real matrices, which can be represented in a way similar to the definition of $\acute{\underline{\Ym}}$ in (\ref{Vstruct}). In order to find the required matrix $\acute{\Vm}$, considering Lemma $1$ of \cite{JD}, the following sequence of real Givens rotations are used as a counterpart of (\ref{GivensVcomplex})
\begin{multline}\label{GivensV}
    \acute{\Vm} \!=\! \! \! \prod_{N_{Sweeps}} \prod_{\substack{1 \leq p,q \leq N_t\\p \neq q}} \! \! \Gfd_{p,q}(\theta) \Gfd_{p+N_t,q+N_t}(\theta) \Gfd_{p,q+N_t}(\dot{\theta}) \\ \Gfd_{q,p+N_t}(\dot{\theta}) \prod_{1 \leq p \leq N_t} {\Gfd_{p,p+N_t}(\ddot{\theta})}
\end{multline}
The rotations $\Gfd_{p,q}(\theta)$ and $\Gfd_{p+N_t,q+N_t}(\theta)$ are applied successively using the same angle parameter $(\theta)$. Similarly, the rotations $\Gfd_{p,q+N_t}(\dot{\theta})$ and $\Gfd_{q,p+N_t}(\dot{\theta})$ are applied with another angle parameter $(\dot{\theta})$. Note that, these rotations are paired in this way to preserve the structure of $\acute{\Vm}$ given in (\ref{Vstruct}) \cite{JD}. The rotation $\Gfd_{p,p+N_t}(\ddot{\theta})$ is applied to deal with the phase shift introduced by the diagonal entries of the mixing matrix $\Am$. The angle parameters $(\theta)$ , $(\dot{\theta})$ and $(\ddot{\theta})$ are computed is such a way to minimize the MM criterion (\ref{MMcost}), using above explained iterative method. For that, we express the MM cost function in terms of the angle parameter $(\theta)$. Now, consider a unitary transformation $\acute{\Zm}=\Gfd_{p,q} \acute{\underline{\Ym}}$, which according to (\ref{GivensMatrix}) only changes the rows `$p$' and `$q$' of $\acute{\underline{\Ym}}$ so that
\begin{equation}\label{GivenPRonY}
    \begin{gathered}
        \acute{z}_{ji}=\acute{\underline{y}}_{ji} ~ \text{for} ~ j \neq p,q, \quad \acute{z}_{pi}=\cos(\theta) \acute{\underline{y}}_{pi} + \sin(\theta) \acute{\underline{y}}_{qi}\\
        \acute{z}_{qi}=-\sin(\theta) \acute{\underline{y}}_{pi} + \cos(\theta) \acute{\underline{y}}_{qi}
    \end{gathered}
\end{equation}
Similarly, the rotation\footnote{For simplicity, we keep the notation $\acute{\underline{\Ym}}$ unchanged even though the matrix is modified after each rotation.} $\Gfd_{p+N_t,q+N_t}$ with the same angle parameter $(\theta)$ modifies the rows `$p+N_t$' and `$q+N_t$' in a similar way as shown in (\ref{GivenPRonY}). Now, (\ref{MMcost}) can be rewritten in terms of $(\theta)$ as (omitting the terms of $\acute{\Zm}$ that are independent of $(\theta)$ and assuming for simplicity that $R_R=R_I=R$)
\begin{multline}\label{GivenCost}
    \Jc_{\text{MMA}}(\theta) = \sum_{i=1}^{N_s} \left[ \left( \acute{z}_{pi}^2-R \right)^2 + \left( \acute{z}_{qi}^2-R \right)^2 \right. \\ \left. + \left( \acute{z}_{p+N_t,i}^2-R \right)^2 + \left( \acute{z}_{q+N_t,i}^2-R \right)^2 \right]
\end{multline}
Using (\ref{GivenPRonY}) and double angle identities we can write
\begin{equation}\label{GivenYpjYqj}
    \acute{z}_{pi}^2 \!=\! \tv_i^\trasp \vv+\frac{1}{2} \left( \acute{\underline{y}}_{pi}^2+\acute{\underline{y}}_{qi}^2 \right), ~ \acute{z}_{qi}^2 \!=\! -\tv_i^\trasp \vv+\frac{1}{2} \left( \acute{\underline{y}}_{pi}^2+\acute{\underline{y}}_{qi}^2 \right)
\end{equation}
where
\begin{equation}\label{vandtj}
    \vv \! \!=\! \! \begin{bmatrix} \cos(2\theta) & \sin(2\theta) \end{bmatrix}^\trasp, ~ \tv_i \! \!=\! \! \begin{bmatrix} \frac{1}{2}(\acute{\underline{y}}_{pi}^2 - \acute{\underline{y}}_{qi}^2) & \acute{\underline{y}}_{pi} \acute{\underline{y}}_{qi} \end{bmatrix}^\trasp
\end{equation}
This allows us to express the first two terms in (\ref{GivenCost}) as
\begin{equation}\label{GivenGpq}
     \! \left( \acute{z}_{pi}^2 \!-\! R \right)^2 \!+\! \left( \acute{z}_{qi}^2 \!-\! R \right)^2 \! \!=\! 2 \vv^\trasp \tv_i \tv_i^\trasp \vv + 2 \left( \frac{\acute{\underline{y}}_{pi}^2 \!+\! \acute{\underline{y}}_{qi}^2}{2} \!-\! R \right)^2
\end{equation}
Similarly, the terms $z_{p+N_t,i}^2$ and $z_{q+N_t,i}^2$ are obtained by replacing $\tv_i$ with $\acute{\tv}_i$ corresponding to indices `$p+N_t$' and `$q+N_t$' in \eqref{vandtj}. Disregarding the constant terms in (\ref{GivenGpq}), we can express $\Jc_{\text{MMA}}(\theta)$ as a quadratic form
\begin{equation}\label{GivenCostPQPMQM}
        \Jc_{\text{MMA}}(\theta) = \vv^\trasp \sum_{i=1}^{N_s}{ \left[ \tv_i \tv_i^\trasp + \acute{\tv}_i {\acute{\tv}_i}^\trasp \right] } \vv = \vv^\trasp \Tm \vv
\end{equation}
The solution $\vv^\circ=\begin{bmatrix} v_1^\circ & v_2^\circ \end{bmatrix}^\transp$ that minimizes (\ref{GivenCostPQPMQM}) is given by the unit norm eigenvector of $\Tm$ corresponding to its smallest eigenvalue, so using (\ref{vandtj}), we can write
\begin{equation}\label{GivenThetaResult}
        \cos(\theta) = \sqrt{\frac{1+v_1^\circ}{2}} \quad \text{and} \quad \sin(\theta) = \frac{v_2^\circ}{\sqrt{2(1+v_1^\circ)}}
\end{equation}
Using (\ref{GivenThetaResult}), the computation of $\Gfd_{p,q}$ and $\Gfd_{p+N_t,q+N_t}$ follows directly from (\ref{GivensMatrix}). Givens rotations $\Gfd_{p,q+N_t}(\dot{\theta})$ and $\Gfd_{q,p+N_t}(\dot{\theta})$ are found similarly and applied successively on $\acute{\underline{\Ym}}$ to compute the filtered separation matrix $\acute{\Vm}$ according to (\ref{GivensV}). The Givens rotation $\Gfd_{p,p+N_t}(\ddot{\theta})$ for `$p=q$' can be similarly found by following the above explained method. By replacing `$q$' with `$p+N_t$' in (\ref{vandtj}) and (\ref{GivenGpq}), the cost function (\ref{MMcost}) (with the constant terms omitted) can be written as
\begin{equation}\label{GivenCostPPM}
        \Jc_{\text{MMA}}(\ddot{\theta}) = \vv^\trasp \sum_{i=1}^{N_s}{ \left[ \tv_i \tv_i^\trasp \right] } \vv = \vv^\trasp \acute{\Tm} \vv
\end{equation}
Hence, the solution $\vv^\circ$ is the least unit norm eigenvector of $\acute{\Tm}$ and $\Gfd_{p,p+N_t}(\ddot{\theta})$ is computed using (\ref{GivenThetaResult}) and (\ref{GivensMatrix}). Matrix $\acute{\Vm}$ is initialized as $\acute{\Vm}=\Id_{2N_t}$ and the overall algorithm is summarized in Table \ref{GMMAAlgo}.

\begin{table}[tb!]
\caption{Givens MMA (G-MMA) Algorithm}
\label{GMMAAlgo}
\begin{boxedalgorithmic}
\STATE Initialization: $\acute{\Vm} = \Id_{2N_t}$
\STATE 1. Pre-whitening: $\underline{\Ym}=\Bm \Ym$ \hfill $\mathcal{O}(N_s N_r^2)$
\STATE 2. Construct real matrix $\acute{\underline{\Ym}}$ using (\ref{Vstruct})
\STATE 3. Givens Rotations: \hfill ($20 N_s N_t^2) + \mathcal{O}(N_s N_t)$/Sweep
\FOR{$n=1:N_{Sweeps}$}
    \FOR{$p=1:N_t$}
        \FOR{$q=p:N_t$}
            \IF{$p=q$}
                \STATE a) Compute $\Gfd_{p,p+N_t}$ using (\ref{GivenCostPPM}), (\ref{GivenThetaResult}) and (\ref{GivensMatrix}) for $\ddot{\theta}$ \hfill ($6N_s$)
                \STATE b) $\acute{\underline{\Ym}} = \Gfd_{p,p+N_t} \acute{\underline{\Ym}}$ \hfill ($4N_s$)
                \STATE c) $\acute{\Vm}=\Gfd_{p,p+N_t} \acute{\Vm}$
            \ELSE
                \STATE d) Compute $\Gfd_{p,q} \, \& \, \Gfd_{p+N_t,q+N_t}$ using (\ref{GivenCostPQPMQM}), (\ref{GivenThetaResult}) and (\ref{GivensMatrix}) for same ($\theta$) \hfill ($12N_s$)
                \STATE e) $\acute{\underline{\Ym}} = \Gfd_{p,q} \, \Gfd_{p+N_t,q+N_t} \acute{\underline{\Ym}}$ \hfill ($8N_s$)
                \STATE f) $\acute{\Vm}=\Gfd_{p,q} \, \Gfd_{p+N_t,q+N_t} \acute{\Vm}$
                \STATE \textbf{repeat} (d \TO f) for $(p,q+N_t) ~\&~ (q,p+N_t)$ using same ($\dot{\theta}$) \hfill ($20N_s$)
            \ENDIF
        \ENDFOR
    \ENDFOR
\ENDFOR
\STATE 4. Estimate the complex sources from $\acute{\underline{\Ym}}$ using \eqref{SMRecInK} and \eqref{Vstruct}.
\end{boxedalgorithmic}
\end{table}

\section{Hyperbolic G-MMA (HG-MMA)}\label{sec:HGMMA}
For a small number of samples $N_s$, the pre-whitening operation is not effective and thus the transformed mixing matrix $\Am$ may be far from unitary. In this case, the performance of G-MMA deteriorates and thus the J-unitary real hyperbolic rotations are applied alternatively along with the Givens rotations to overcome this limitation. This results in an algorithm named Hyperbolic Givens MMA (HG-MMA). So, now the matrix $\acute{\Vm}$ can be decomposed into a product of elementary hyperbolic rotations $\Hfd_{p,q}$, Givens rotations $\Gfd_{p,q}$, and normalization transformation $\Nfd_{p,q}$ as follows
\begin{multline}\label{HyperGivensV}
     \acute{\Vm} = \prod_{N_{Sweeps}} \prod_{\substack{1 \leq p,q \leq N_t\\p \neq q}} \Gammam_{p,q}(\theta, \gamma) \Gammam_{p+N_t,q+N_t}(\theta, \gamma)\\ \Gammam_{p,q+N_t}(\dot{\theta}, \dot{\gamma}) \Gammam_{q,p+N_t}(\dot{\theta}, -\dot{\gamma}) \prod_{1 \leq p \leq N_t}{\Gfd_{p,p+N_t}(\ddot{\theta})}
\end{multline}
where $\Gammam_{p,q}=\Nfd_{p,q} \Gfd_{p,q} \Hfd_{p,q}$. Similar to the Givens rotations, the hyperbolic rotations $\Hfd_{p,q}$ and $\Hfd_{p+N_t,q+N_t}$ are applied using the same parameter ($\gamma$) while $\Hfd_{p,q+N_t}$ and $\Hfd_{q,p+N_t}$ are applied using another same but opposite parameter ($\dot{\gamma}$) and ($-\dot{\gamma}$), respectively. We will consider dispersion parameters $R_R$ and $R_I$ be equal to $1$ and use $\Nfd_{p,q}$ for normalization. Below we give a brief of finding the hyperbolic and the normalization transformation parameters to minimize the MM criterion (\ref{MMcost}).

\subsection{Computation of Hyperbolic and Givens rotations}\label{sec:Hyp_Giv_rot_cal}
Let us consider one hyperbolic transformation $\acute{\Zm} = \Hfd_{p,q} \acute{\underline{\Ym}}$, which modifies $\acute{\underline{\Ym}}$ according to
\begin{equation}\label{HyperPRonY}
    \begin{gathered}
        \acute{z}_{ji} \!=\! \acute{\underline{y}}_{ji} ~ \text{for} ~ j \neq p,q, \quad\acute{z}_{pi} \!=\! \cosh(\gamma) \acute{\underline{y}}_{pi} + \sinh(\gamma) \acute{\underline{y}}_{qi}\\
        \acute{z}_{qi}=\sinh(\gamma) \acute{\underline{y}}_{pi} + \cosh(\gamma) \acute{\underline{y}}_{qi}
    \end{gathered}
\end{equation}
Now, using hyperbolic double angle identities we obtain
\begin{equation}\label{HyperYpjYqj}
        \acute{z}_{pi}^2 \!=\! \rv_i^\trasp \uv+\frac{1}{2} \left( \acute{\underline{y}}_{pi}^2-\acute{\underline{y}}_{qi}^2 \right), ~
        \acute{z}_{qi}^2 \!=\! \rv_i^\trasp \uv-\frac{1}{2} \left( \acute{\underline{y}}_{pi}^2-\acute{\underline{y}}_{qi}^2 \right)
\end{equation}
where
\begin{equation}\label{uandrj}
    \uv \! \!=\! \! \begin{bmatrix} \cosh(2\gamma) & \! \! \sinh(2\gamma) \end{bmatrix}^\trasp, ~
    \rv_i \! \!=\! \! \begin{bmatrix} \frac{1}{2}(\acute{\underline{y}}_{pi}^2 + \acute{\underline{y}}_{qi}^2) & \! \! \acute{\underline{y}}_{pi} \acute{\underline{y}}_{qi} \end{bmatrix}^\trasp \! \! \!
\end{equation}
Similar expressions can be derived for $z_{p+N_t,i}^2$ and $z_{q+N_t,i}^2$. Substituting these expressions in (\ref{GivenCost}) and omitting the terms that are independent of ($\gamma$) yields
\begin{equation}\label{HyperCostFinal}
    \begin{aligned}
        \Jc_{\text{MMA}}(\gamma)& \!=\! \uv^\trasp \! \! \left[ \sum_{i=1}^{N_s}{\rv_i \rv_i^\trasp + \acute{\rv}_i \acute{\rv}_i^\trasp} \right] \! \! \uv - 2 \uv^\trasp \! \! \left[ \sum_{i=1}^{N_s}{\rv_i + \acute{\rv}_i} \right] \! \! \\
        &= \uv^\trasp \Rm \uv - 2 \uv^\trasp \rv
    \end{aligned}
\end{equation}
where $\acute{\rv}_i = \begin{bmatrix} \frac{1}{2}(\acute{\underline{y}}_{p+N_t,i}^2 + \acute{\underline{y}}_{q+N_t,i}^2) & \acute{\underline{y}}_{p+N_t,i} \acute{\underline{y}}_{q+N_t,i} \end{bmatrix}^\trasp$. The optimization problem in (\ref{HyperCostFinal}) can be solved using either Lagrange multiplier method (exact solution) or by taking linear approximation of hyperbolic sine and cosine around zero (approximate solution). Both methods are discussed below.

\subsubsection{Exact Solution}\label{Exact_Sol_HGMMA}
We consider the constrained optimization
\begin{equation}\label{HyperLagrange}
    \min_\uv ~~ \Fc(\uv) = \uv^\trasp \Rm \uv - 2 \rv^\trasp \uv ~~~~ \text{s.t.} ~~ \uv^\trasp \Jm_2 \uv=1
\end{equation}
where $\Jm_2=\diag\left(\begin{bmatrix}1 & -1 \end{bmatrix}\right)$ corresponding to $\cosh^2(2\gamma)-\sinh^2(2\gamma)=1$. The Lagrangian of (\ref{HyperLagrange}) can be written as
\begin{equation}\label{HyperLagrangian}
    \Lc(\uv,\lambda) = \uv^\trasp \Rm \uv - 2 \rv^\trasp \uv + \lambda \left( \uv^\trasp \Jm_2 \uv - 1 \right)
\end{equation}
The solution of this Lagrangian is given by
\begin{equation}\label{HyperLagrangeSol}
    \uv=(\Rm+\lambda \Jm_2)^{-1} \rv
\end{equation}
Using (\ref{HyperLagrangeSol}), the constraint equation results in a $4^{\text{th}}$ order polynomial equation
\begin{equation}\label{LagLambdaSol}
    \rv^\trasp (\Rm+\lambda \Jm_2)^{-1} \Jm_2 (\Rm+\lambda \Jm_2)^{-1} \rv =1
\end{equation}
Of the four roots of (\ref{LagLambdaSol}), we use the real value\footnote{In the case the set of solutions is empty, we set by default $\lambda=0$.} of $\lambda$ that results in the minimum value of $\Lc(\uv,\lambda)$ with a vector $\uv$ satisfying $u_1>0$. We then solve for $\uv^\circ=[u_1^\circ, u_2^\circ]^\trasp$ from (\ref{HyperLagrangeSol}) and solve for the hyperbolic sine and cosine of $(\gamma)$ as
\begin{equation}\label{HyperPar}
    \cosh(\gamma)=\sqrt{ \frac{1+u_1^\circ}{2} } ~~and~~ \sinh(\gamma)=\frac{u_2^\circ}{\sqrt{2(1+u_1^\circ)}}
\end{equation}
which allows us to construct the hyperbolic rotations $\Hfd_{p,q}$ and $\Hfd_{p+N_t,q+N_t}$ defined in (\ref{HyperMatrix}).

For the remaining hyperbolic rotations $\Hfd_{p,q+N_t}$ and $\Hfd_{q,p+N_t}$, the optimization problem in (\ref{HyperCostFinal}) is conducted for the other hyperbolic parameter $(\dot{\gamma})$, where $\rv_i = \begin{bmatrix} \frac{1}{2}(\acute{\underline{y}}_{pi}^2 + \acute{\underline{y}}_{q+N_t,i}^2) & \acute{\underline{y}}_{pi} \acute{\underline{y}}_{q+N_t,i} \end{bmatrix}^\trasp$ and $\acute{\rv}_i = \begin{bmatrix} \frac{1}{2}(\acute{\underline{y}}_{qi}^2 + \acute{\underline{y}}_{p+N_t,i}^2) & -\acute{\underline{y}}_{qi} \acute{\underline{y}}_{p+N_t,i} \end{bmatrix}^\trasp$. Then, the modified optimization problem is minimized using the same method as explained above. This provides the solution $\acute{\uv^\circ}=[\acute{u}_1^\circ, \acute{u}_2^\circ]^\trasp$ and the hyperbolic angles are obtained using \eqref{HyperPar} for hyperbolic parameter $(\dot{\gamma})$. The computation of the hyperbolic rotations $\Hfd_{p,q+N_t}(\dot{\gamma})$ and $\Hfd_{q,p+N_t}(-\dot{\gamma})$ follows directly from \eqref{HyperPar} and (\ref{HyperMatrix}). Note that these rotations are applied using same but opposite hyperbolic angle parameter ($\dot{\gamma}$).

\subsubsection{Approximate Solution}\label{Approx_Sol_HGMMA}
In this approach, we will consider the linear approximation of hyperbolic sine and cosine around zero given by $\sinh(2 \gamma) \approx 2 \sinh(\gamma)$ and $\cosh(2 \gamma) \approx \cosh(\gamma)$. Now, let us define the elements of symmetric matrix $\Rm$ and vector $\rv$ used in (\ref{HyperCostFinal}) as
\begin{equation}\label{MatRandVecr}
    \Rm = \begin{bmatrix} r_{11} & r_{12} \\ r_{21} & r_{22} \end{bmatrix} ~~\text{and}~~ \rv = \begin{bmatrix} r_{1} \\ r_{2} \end{bmatrix}
\end{equation}
Using (\ref{uandrj}), (\ref{MatRandVecr}) and neglecting the terms independent of $(\gamma)$, the cost function (\ref{HyperCostFinal}) can be rewritten as
\begin{multline}\label{HyperCostAppr}
    \Jc_{\text{MMA}}(\gamma) = \cosh(4 \gamma) \frac{r_{11} + r_{22}}{2} + \sinh(4 \gamma) r_{12} \\ - 2 \cosh(2 \gamma) r_1 - 2 \sinh(2 \gamma) r_2
\end{multline}
Setting the derivative of (\ref{HyperCostAppr}) w.r.t $(\gamma)$ to zero and using the previous approximation, we obtain
\begin{equation}\label{HyperCostApprFinal}
    \sinh(2 \gamma) \left( r_{11} + r_{22} - r_1 \right) - \cosh(2 \gamma) \left( r_2 - r_{12} \right) = 0
\end{equation}
and thus the solution $(\gamma)$ is
\begin{equation}\label{HyperSolGamma}
    \gamma = \frac{1}{2} \arctanh{ \left( \frac{ r_2 - r_{12} }{ r_{11} + r_{22} - r_1 } \right) }
\end{equation}
In a similar way, the hyperbolic rotation parameter ($\dot{\gamma}$) can be found using appropriate $\Rm$ and $\rv$ as explained in section \ref{Exact_Sol_HGMMA}. The hyperbolic rotations are computed using (\ref{HyperSolGamma}) and (\ref{HyperMatrix}) and applied accordingly as explained in section \ref{Exact_Sol_HGMMA}. After applying the hyperbolic rotations, Givens rotations are applied in a similar way as explained in section \ref{sec:GMMA} and then normalization rotations are applied as explained below.

\subsection{Calculating the normalization transformations}\label{sec:Norm_Cal_HGMMA}
The normalization is applied to compensate for the dispersion parameters $R_R$ and $R_I$. Let's consider that we have transformed only one row `$p$' of matrix $\underline{\Ym}$, which corresponds to the transformation of rows `$p$' and `$p+N_t$' for matrix $\acute{\underline{\Ym}}$. In this case, the normalization transformation $\Nfd_{p}$ is an identity matrix except for the two diagonal elements $\Nc_{pp}=\Nc_{p+N_t,p+N_t}=\lambda_p$ and the MM cost function (\ref{MMcost}) (with the constant terms omitted) becomes
\begin{equation}\label{NormMMCost}
    \! ~ \Jc_{\text{MMA}}(\lambda_p) \! \!=\! \! \sum_{i=1}^{N_s}{ \! \left[ \! \left( \! \! \left( \lambda_p \acute{\underline{y}}_{pi} \! \right)^2 \! \! - 1 \right)^2 \! \! + \! \! \left( \! \! \left( \lambda_p \acute{\underline{y}}_{p+N_t,i} \right)^2 \! \! - 1 \right)^2 \right] } \! \! \! \! \! \! \! \!
\end{equation}
Taking the derivative of (\ref{NormMMCost}) w.r.t $(\lambda_p)$ and setting the result to zero gives optimal normalization parameter
\begin{equation}\label{NormPar}
    \lambda_p = \sqrt{ \frac{ \sum_{i=1}^{N_s}{ \acute{\underline{y}}_{pi}^2 + \acute{\underline{y}}_{p+N_t,i}^2 } }{ \sum_{i=1}^{N_s}{ \acute{\underline{y}}_{pi}^4 + \acute{\underline{y}}_{p+N_t,i}^4 } } }, ~~~\forall ~p
\end{equation}

In our simulations, we observed that the normalization rotation is not necessary at each step and can be performed only once per sweep. In this case, the diagonal entries of matrix $\Nfd$ are $\Nc_{pp}=\Nc_{p+N_t,p+N_t}=\lambda_p$ given as in (\ref{NormPar}) where $1 \leq p \leq N_t$. HG-MMA is presented in Table \ref{HGMMAAlgo}.

\begin{table}[tb!]
\caption{Hyperbolic Givens MMA (HG-MMA) Algorithm}
\label{HGMMAAlgo}
\begin{boxedalgorithmic}
\STATE Initialization: $\acute{\Vm} = \Id_{2N_t}$
\STATE Subspace projection or approximate pre-whitening if $N_r>N_t$ \hfill $\mathcal{O}(N_s N_t^2)$
\STATE 1. Create real matrix $\acute{\underline{\Ym}}$ using (\ref{Vstruct})
\STATE 2. Hyperbolic, Givens $\&$ Normalization Rotations: \hfill $(40 N_s N_t^2)+\mathcal{O}(N_s N_t)$
\FOR{$n=1:N_{Sweeps}$}
    \FOR{$p=1:N_t$}
        \FOR{$q=p:N_t$}
            \IF{$p=q$}
                \STATE a) Apply Givens rotation using (a \TO c) of Table \ref{GMMAAlgo} \hfill ($10N_s$)
            \ELSE
                \STATE b) Compute $\Hfd_{p,q} ~\&~ \Hfd_{p+N_t,q+N_t}$ using (\ref{HyperPar}) and (\ref{HyperMatrix}) for ($\gamma$) \hfill ($12N_s$)
                \STATE c) $\acute{\underline{\Ym}} = \Hfd_{p,q} \Hfd_{p+N_t,q+N_t} \acute{\underline{\Ym}}$ \hfill ($8N_s$)
                \STATE d) $\acute{\Vm}=\Hfd_{p,q} \Hfd_{p+N_t,q+N_t} \acute{\Vm}$
                \STATE e) Apply Givens rotation using (d \TO f) of Table \ref{GMMAAlgo} \hfill ($20N_s$)
                \STATE \textbf{repeat} steps (b \TO e) for $(p,q+N_t) ~\&~ (q,p+N_t)$ using $(\dot{\theta},\dot{\gamma}) ~\&~ (\dot{\theta},-\dot{\gamma})$, respectively  \hfill ($40N_s$)
            \ENDIF
        \ENDFOR
    \ENDFOR
    \STATE f) Compute $\Nfd$ using (\ref{NormPar}) \hfill ($6N_s N_t$)
    \STATE g) $\acute{\underline{\Ym}} = \Nfd \acute{\underline{\Ym}}$ \hfill ($2 N_s N_t$)
    \STATE h) $\acute{\Vm}=\Nfd \acute{\Vm}$
\ENDFOR
\end{boxedalgorithmic}
\end{table}

\section{Givens AMA (G-AMA)}\label{sec:GAMA_Algo}
For the design of AM algorithms, the AM cost function \eqref{AMA_Cost} is selected because of the following reasons: i) it satisfies all three properties presented in Section \ref{AM_Cost_Func}, which are sufficient conditions to shape the cost function for high-order square QAM signals, ii) it is the simplest among all other AM cost functions and computationally less expensive. Note that, the number of computations in this one is independent of the number of alphabet points as opposed to AM cost functions in \cite{Barbarossa1997, Li1995}, iii) it deals with the real and imaginary parts of the output signal, separately. Thus, it is relatively easier to optimize using real Givens and hyperbolic rotations. In this section, G-MMA is used as an initialization followed by optimization of AM cost function (with $n=1$ for CME term in \eqref{CMETerm}) using real Givens and hyperbolic rotations, which results in algorithms G-AMA and HG-AMA. The combination of MMA and AMA is not new and recently used by Labed \emph{et al.} \cite{labed2013} for the problem of blind equalization.

After using G-MMA for the initialization, the matrix $\acute{\Vm}$ is updated using the following Givens rotations
\begin{multline}\label{GivensV_AMA}
    \acute{\Vm}^{n} = \prod_{\substack{1 \leq p,q \leq N_t\\p \neq q}} \Gfd_{p,q+N_t}(\dot{\theta}) \Gfd_{q,p+N_t}(\dot{\theta}) \Gfd_{p,q}(\theta) \\ \Gfd_{p+N_t,q+N_t}(\theta) \acute{\Vm}^{n-1}
\end{multline}
where $n=n_0+1,\ldots,N_{Sweeps}$, where $N_{Sweeps}$ is the number of iterations of G-AMA until convergence and $n_0$ is the number of iterations of G-MMA for initialization\footnote{As per observations from the rate of convergence for G-MMA, it converges in $n_0 = 5$ for the considered cases. However, one can choose the number of sweeps as the one corresponding to an almost flat variation of $\Jc_{\text{MMA}}$.}. Let us express the AM cost function in terms of the angle parameter ($\theta$) which is computed such that $\Jc_{\text{AMA}}(\theta)$ is minimized. Using similar derivations as before, one can write
\begin{equation}\label{GivenCost_AMA}
    \! ~ \Jc_{\text{AMA}} \! \!=\! \! \sum_{i=1}^{N_s} \left[ g\left(\acute{z}_{pi}\right) + g\left(\acute{z}_{qi}\right) + g\left(\acute{z}_{p+N_t,i}\right) + g\left(\acute{z}_{q+N_t,i}\right) \right] \! \! \! \!
\end{equation}
where the first two terms in \eqref{GivenCost_AMA} can be defined with $n=1$ in \eqref{CMETerm}  as
\begin{equation}\label{Givens_AMA_Terms}
  \begin{aligned}
    g\left(\acute{z}_{pi}\right) &= 1 - \sin^{2}\left\{ \left( \cos(\theta) \acute{\underline{y}}_{pi} + \sin(\theta) \acute{\underline{y}}_{qi} \right) \left( \frac{\pi}{2d} \right) \right\} \\
    g\left(\acute{z}_{qi}\right) &= 1 - \sin^{2}\left\{ \left( -\sin(\theta) \acute{\underline{y}}_{pi} + \cos(\theta) \acute{\underline{y}}_{qi} \right) \left( \frac{\pi}{2d} \right) \right\} \! \!
  \end{aligned}
\end{equation}
and the last two terms are obtained by replacing `$p$' and `$q$' with `$p+N_t$' and `$q+N_t$' in \eqref{Givens_AMA_Terms}, respectively. The bounded non-linear optimization problem can now be stated as
\begin{equation}\label{OptimCost}
    \min_\theta ~~ \Jc_{\text{AMA}} ~~~~ \text{s.t.} ~~ \theta \in \left[ -\pi/4, \pi/4 \right]
\end{equation}
The optimization problem in \eqref{OptimCost} can be solved either by using MATLAB optimization toolbox that can be termed as `exact solution' or by using Taylor series approximation of trigonometric functions around zero, which will be referred to as `approximate solution'. This approximation can be justified using Figure \ref{OrigAMACost}, which plots the values of AMA cost function $\Jc_{\text{AMA}}$ in \eqref{GivenCost_AMA} vs. $\theta$ for some random received pre-whitened signal $\acute{\underline{\Ym}}$ after 5 sweeps of G-MMA with $N_t = 3, N_r = 5, N_s = 300$, SNR $= 30$dB and normalized 64-QAM constellation. Note that the optimum $\theta^\circ$ is very close to zero. Thus in the following section, we show that for a certain range of $\theta$ close to zero, the approximation fits very well with the original values of the AMA cost function.

\subsection{Exact Solution}\label{exact_sol_GAMA}
For the exact solution, the objective function in \eqref{GivenCost_AMA} is passed to the MATLAB optimization toolbox\footnote{Our objective here is just to compute an `exact' solution of \eqref{OptimCost}, which can be obtained by a linesearch algorithm as well.} `{\bf fminsearchbnd}' along with $\theta_0 = 0.001$ as a starting point and bounds $\theta \in \left[ -\pi/4, \pi/4 \right]$, in order to find optimum $\theta^\circ$ for the minimization of \eqref{OptimCost}. Givens rotation matrices $\Gfd_{p,q}(\theta^\circ)$ and $\Gfd_{p+N_t,q+N_t}(\theta^\circ)$ are then applied to update $\acute{\Vm}$ according to \eqref{GivensV_AMA}. The remaining Givens rotations $\Gfd_{p,q+N_t}(\dot{\theta})$ and $\Gfd_{q,p+N_t}(\dot{\theta})$ can be found similarly by replacing subscripts accordingly in \eqref{GivenCost_AMA} and \eqref{Givens_AMA_Terms} and then computing optimum $\dot{\theta}^\circ$. Then, the separation matrix $\acute{\Vm}$ is updated again according to \eqref{GivensV_AMA}. This process is repeated until convergence.

\subsection{Approximate Solution}\label{approx_sol_GAMA}
\begin{figure}[tb!]\centering
    \setlength\figureheight{3.5cm}
	\setlength\figurewidth{8.2cm}
    \input{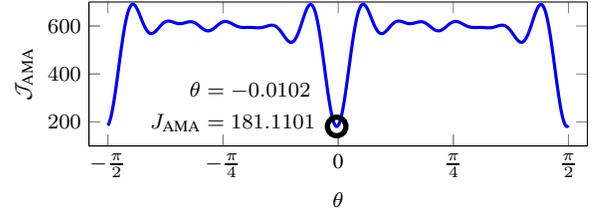}
    \caption{$\Jc_{\text{AMA}}$ vs. $\theta$ for random received pre-whitened signal after 5 sweeps of G-MMA with $N_t = 3, N_r = 5, N_s = 300$, SNR $= 30$dB and normalized 64-QAM constellation.}
    \label{OrigAMACost}
\end{figure}

As observed from Figure \ref{OrigAMACost}, the optimum $\theta^\circ$ is very close to zero, thus the Taylor series approximation around zero can be applied. Here, we will consider the approximation up to the $4^{\text{th}}$ order using the following approximate identities
\begin{equation}\label{ApproxCosSin}
    \sin(\theta) \approx \theta - \frac{\theta^3}{6}, \qquad \cos(\theta) \approx 1 - \frac{\theta^2}{2} + \frac{\theta^4}{24}
\end{equation}
Now, using the approximation in \eqref{ApproxCosSin} to `$\cos(\theta)$' and `$\sin(\theta)$' in \eqref{Givens_AMA_Terms} and expanding the terms, results in
\begin{equation}\label{gZpAppr2}
    \! ~ g\left(\acute{z}_{pi}\right) \! \! \approx \! \! \frac{1}{48 d^4} c_4^{pi} \theta^4 +\! \frac{1}{12 d^3} c_3^{pi} \theta^3 +\! \frac{1}{4 d^2} c_2^{pi} \theta^2 -\! \frac{1}{2d} c_1^{pi} \theta +\! \frac{1}{2} c_0^{pi} \! \! \! \!
\end{equation}
where
\begin{equation}\label{coeff}
    \begin{aligned}
        c_4^{pi} &= 4 \pi^2 d^2 \acute{\underline{y}}_{qi}^2 \cos \left(\frac{\pi  \acute{\underline{y}}_{pi}}{d}\right) + \pi^4 \acute{\underline{y}}_{qi}^4 \cos \left(\frac{\pi \acute{\underline{y}}_{pi}}{d}\right)\\
         &~~ - 3 \pi^2 d^2 \acute{\underline{y}}_{pi}^2 \cos \left(\frac{\pi \acute{\underline{y}}_{pi}}{d}\right) - \pi d^3 \acute{\underline{y}}_{pi} \sin \left(\frac{\pi  \acute{\underline{y}}_{pi}}{d}\right)\\
         &~~ - 6 \pi^3 d \acute{\underline{y}}_{pi} \acute{\underline{y}}_{qi}^2 \sin \left(\frac{\pi  \acute{\underline{y}}_{pi}}{d}\right)\\
        c_3^{pi} &= \pi  d^2 \acute{\underline{y}}_{qi} \sin \left(\frac{\pi  \acute{\underline{y}}_{pi}}{d}\right) + \pi ^3 \acute{\underline{y}}_{qi}^3 \sin \left(\frac{\pi  \acute{\underline{y}}_{pi}}{d}\right) \\
        &~~ +3 \pi ^2 d \acute{\underline{y}}_{pi} \acute{\underline{y}}_{qi} \cos\left(\frac{\pi  \acute{\underline{y}}_{pi}}{d}\right) \\
        c_2^{pi} &= \pi  d \acute{\underline{y}}_{pi} \sin \left(\frac{\pi  \acute{\underline{y}}_{pi}}{d}\right) - \pi ^2 \acute{\underline{y}}_{qi}^2 \cos \left(\frac{\pi  \acute{\underline{y}}_{pi}}{d}\right) \\
        c_1^{pi} & = \pi  \acute{\underline{y}}_{qi} \sin \left(\frac{\pi  \acute{\underline{y}}_{pi}}{d}\right), \quad c_0^{pi} = 1 + \cos \left(\frac{\pi  \acute{\underline{y}}_{pi}}{d}\right)
    \end{aligned}
\end{equation}
Similarly $g\left(\acute{z}_{qi}\right)$ can be approximated by
\begin{equation}\label{gZq}
    \! ~ g\left(\acute{z}_{qi}\right) \! \! \approx \! \! \frac{1}{48 d^4} c_4^{qi} \theta^4 -\! \frac{1}{12 d^3} c_3^{qi} \theta^3 +\! \frac{1}{4 d^2} c_2^{qi} \theta^2 +\! \frac{1}{2d} c_1^{qi} \theta +\! \frac{1}{2} c_0^{qi} \! \! \! \!
\end{equation}
where all the coefficients are obtained by replacing `$p$' with `$q$' and `$q$' with `$p$' in \eqref{coeff}. The $3^{\text{rd}}$ term $g\left(\acute{z}_{p+N_t,i}\right)$ of \eqref{GivenCost_AMA} has the same approximation as given in \eqref{gZpAppr2}, where the coefficients are obtained by replacing `$p$' with `$p+N_t$' and `$q$' with `$q+N_t$' in \eqref{coeff}. The last term $g\left(\acute{z}_{q+N_t,i}\right)$ of \eqref{GivenCost_AMA} is approximated as \eqref{gZq}, where the coefficients are obtained by replacing `$p$' with `$q+N_t$' and `$q$' with `$p+N_t$' in \eqref{coeff}.

Now, using \eqref{gZpAppr2} and \eqref{gZq} in cost function \eqref{GivenCost_AMA} results in the $4^{\text{th}}$ order polynomial equation
\begin{equation}\label{MoenCostPolyAppr}
    \! ~ \Jc_{\text{AMA}} \! \! \approx \! \! \frac{1}{48 d^4} C_4 \theta^4 \!+\! \frac{1}{12 d^3} C_3 \theta^3 \!+\! \frac{1}{4 d^2} C_2 \theta^2 \!+\! \frac{1}{2d} C_1 \theta \!+\! \frac{1}{2} C_0 \! \! \! \! \!
\end{equation}
where the coefficients in \eqref{MoenCostPolyAppr} are given by
\begin{equation}\label{Coef_sum}
  \begin{aligned}
    \! C_l \!&=\! \sum_{i=1}^{N_s} \!\! \left( c_l^{pi} + c_l^{qi} + c_l^{p+N_t,i} + c_l^{q+N_t,i} \right), ~ l \in \{ 0,2,4 \} \\
    \! C_3 \!&=\! \sum_{i=1}^{N_s} \!\! \left( c_3^{pi} - c_3^{qi} + c_3^{p+N_t,i} - c_3^{q+N_t,i} \right) \\
    \! C_1 \!&=\! \sum_{i=1}^{N_s} \!\! \left( - c_1^{pi} + c_1^{qi} - c_1^{p+N_t,i} + c_1^{q+N_t,i} \right) \\ \! \! \!
  \end{aligned}
\end{equation}
Taking the gradient of \eqref{MoenCostPolyAppr} w.r.t. $\theta$ yields
\begin{equation}\label{GAMA_Sol}
  \frac{\partial \Jc_{\text{AMA}}(\theta)}{\partial \theta} \approx \frac{1}{12 d^4} C_4 \theta^3 + \frac{1}{4 d^3} C_3 \theta^2 + \frac{1}{2 d^2} C_2 \theta + \frac{1}{2d} C_1
\end{equation}
where the coefficients are the same as defined in \eqref{Coef_sum}. Out of the three possible roots of \eqref{GAMA_Sol}, the optimum $\theta^\circ$ is selected which results in minimum value of $\Jc_{\text{AMA}}(\theta)$ in \eqref{GivenCost_AMA}.

\begin{figure}[tb!]\centering
    \setlength\figureheight{3.5cm}
	\setlength\figurewidth{8.2cm}
    \input{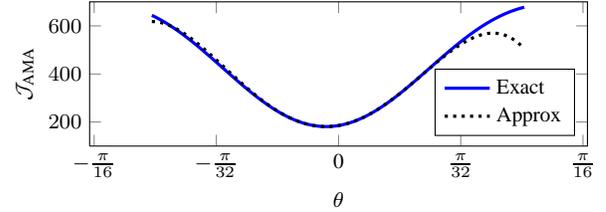}
    \caption{Comparison of exact and approximated G-AMA cost functions after 5 sweeps of G-MMA with $N_t = 3, N_r = 5, N_s = 300$, SNR $= 30$dB and normalized 64-QAM constellation}
    \label{CompCost}
\end{figure}

To illustrate that the approximation in \eqref{MoenCostPolyAppr} is good enough, we have compared the original cost function and approximated one for a certain range of $\theta$ around zero in Figure \ref{CompCost}.

Rotations $\Gfd_{p,q+N_t}(\dot{\theta})$ and $\Gfd_{q,p+N_t}(\dot{\theta})$ are similarly found by replacing subscripts accordingly and computing optimum $\dot{\theta}^\circ$. Then, the rotations are applied successively on $\acute{\underline{\Ym}}$.

In summary, matrix $\acute{\Vm}$ is initialized as identity matrix, then G-MMA is applied for $n_0=5$ followed by the update of matrix $\acute{\Vm}$ according to \eqref{GivensV_AMA} by applying Givens rotations on $\acute{\underline{\Ym}}$ using the above method, until convergence. The overall algorithm is summarized in Table \ref{GAMAAlgo}.

\begin{table}[tb!]
\caption{Givens AMA (G-AMA) Algorithm}
\label{GAMAAlgo}
\begin{boxedalgorithmic}
\STATE Initialization: $\acute{\Vm} = \Id_{2N_t}$
\STATE 1. Pre-whitening: $\underline{\Ym}=\Bm \Ym$
\STATE 2. Construct real matrix $\acute{\underline{\Ym}}$ using (\ref{Vstruct})
\STATE 3. Givens Rotations:
\FOR{$n=1:N_{Sweeps}$}
    \IF{$n<=n_0$}
        \STATE a) Apply G-MMA as given in Table \ref{GMMAAlgo}
    \ELSE
        \FOR{$p=1:N_t-1$}
            \FOR{$q=p+1:N_t$}
                \STATE b) Find optimum ($\theta^\circ$) using roots of \eqref{GAMA_Sol} which gives minimum value of \eqref{GivenCost_AMA}
                \STATE c) Compute $\Gfd_{p,q} \, \& \, \Gfd_{p+N_t,q+N_t}$ using \eqref{GivensMatrix} for same ($\theta^\circ$)
                \STATE d) $\acute{\underline{\Ym}} = \Gfd_{p,q} \, \Gfd_{p+N_t,q+N_t} \acute{\underline{\Ym}}$
                \STATE e) $\acute{\Vm}=\Gfd_{p,q} \, \Gfd_{p+N_t,q+N_t} \acute{\Vm}$
                \STATE \textbf{repeat} (b \TO e) for $(p,q+N_t) ~\&~ (q,p+N_t)$ using same ($\dot{\theta}^\circ$)
            \ENDFOR
        \ENDFOR
    \ENDIF
\ENDFOR
\end{boxedalgorithmic}
\end{table}

\section{Hyperbolic G-AMA (HG-AMA)}\label{sec:HGAMA_Algo}
As stated earlier, for a small number of samples $N_s$, J-unitary real hyperbolic rotations are applied alternatively along with the Givens rotations to overcome the limitation of ill-whitening. This results in an algorithm named as Hyperbolic Givens AMA (HG-AMA).

For HG-AMA, first of all G-MMA is used for initialization. Then, matrix $\Vm$ is updated iteratively until convergence using following hyperbolic $\Hfd_{p,q}$ and Givens $\Gfd_{p,q}$ rotations
\begin{multline}\label{HGV}
        \acute{\Vm}^{n} = \prod_{\substack{1 \leq p,q \leq N_t\\p \neq q}} \Gammam_{p,q+N_t}(\dot{\theta}, \dot{\gamma}) \Gammam_{q,p+N_t}(\dot{\theta}, -\dot{\gamma})\\ \Gammam_{p,q}(\theta, \gamma) \Gammam_{p+N_t,q+N_t}(\theta, \gamma) \acute{\Vm}^{n-1}
\end{multline}
where $\Gammam_{p,q}=\Gfd_{p,q} \Hfd_{p,q}$. Let us express the AM cost function in terms of parameter ($\gamma$) which is computed such that $\Jc_{\text{AMA}}(\gamma)$ is minimized. Now, using similar derivations as before, one can write
\begin{equation}\label{MoenCostHyper}
    \! ~ \Jc_{\text{AMA}} \! \!= \! \! \sum_{i=1}^{N_s} \left[ g\left(\acute{z}_{pi}\right) + g\left(\acute{z}_{qi}\right) + g\left(\acute{z}_{p+N_t,i}\right) + g\left(\acute{z}_{q+N_t,i}\right) \right] \! \! \! \! \!
\end{equation}
where the first two terms in \eqref{MoenCostHyper} can be defined as
\begin{equation}\label{Hyper_AMA_Terms}
  \begin{aligned}
    \! \! \! g\left(\acute{z}_{pi}\right) &= 1 - \sin^{2}\left\{ \left( \cosh(\gamma) \acute{\underline{y}}_{pi} + \sinh(\gamma) \acute{\underline{y}}_{qi} \right) \left( \frac{\pi}{2d} \right) \right\} \! \! \! \\
    \! \! \! g\left(\acute{z}_{qi}\right) &= 1 - \sin^{2}\left\{ \left( \sinh(\gamma) \acute{\underline{y}}_{pi} + \cosh(\gamma) \acute{\underline{y}}_{qi} \right) \left( \frac{\pi}{2d} \right) \right\} \! \! \!
  \end{aligned}
\end{equation}
and the last two terms are obtained by replacing `$p$' and `$q$' with `$p+N_t$' and `$q+N_t$' in \eqref{Hyper_AMA_Terms}, respectively.

\begin{figure}[tb!]\centering
    \setlength\figureheight{3.5cm}
	\setlength\figurewidth{8.2cm}
    \input{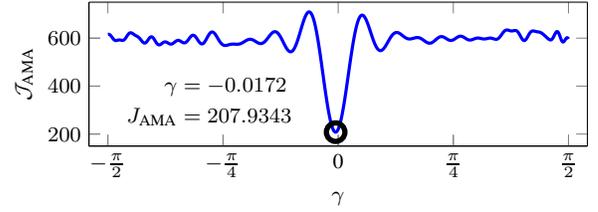}
    \caption{$\Jc_{\text{AMA}}$ vs. $\gamma$ for random received pre-whitened signal after $N_{Sweeps} = 5$ of G-MMA with $N_t = 3, N_r = 5, N_s = 300$, SNR $= 30$dB and normalized 64-QAM constellation.}
    \label{OrigAMACostHyper}
\end{figure}

Figure \ref{OrigAMACostHyper} represents $\Jc_{\text{AMA}}$ in \eqref{MoenCostHyper} vs. $(\gamma)$ after 5 sweeps of G-MMA with $N_t=3, N_r=5, N_s=300$, SNR $= 30$dB and normalized 64-QAM constellation. It can be noticed that optimum $(\gamma^\circ)$ is very close to zero. Thus, we can apply Taylor series approximation of hyperbolic and trigonometric functions around zero, in order to find the solution of the optimization problem in \eqref{MoenCostHyper}. Next, two possible ways are detailed to solve this optimization problem.

\subsection{Exact Solution}\label{sec:Exact_Sol_Hyper}
For the exact solution, the objective function \eqref{MoenCostHyper} is passed to the toolbox `{\bf fminsearch}' with $\gamma_0=0.001$ as starting point which minimizes \eqref{MoenCostHyper} and returns the optimum hyperbolic rotation parameter $(\gamma^\circ)$. $\Hfd_{p,q}(\gamma^\circ)$ and $\Hfd_{p+N_t,q+N_t}(\gamma^\circ)$ are then computed and applied to update $\acute{\Vm}$ according to \eqref{HGV}.

For rotations $\Hfd_{p,q+N_t}(\dot{\gamma})$ and $\Hfd_{q,p+N_t}(-\dot{\gamma})$, $1^{\text{st}}$ and $4^{\text{th}}$ terms of the objective function in \eqref{MoenCostHyper} are defined as
\begin{equation}\label{Hyper_AMA_Terms2}
  \begin{aligned}
    \! \! \! \! \! \! g\left(\acute{z}_{pi}\right) \! \! &= \! \! \cos^{2}\left\{ \! \left( \cosh(\dot{\gamma}) \acute{\underline{y}}_{pi} + \sinh(\dot{\gamma}) \acute{\underline{y}}_{q+N_t,i} \right) \! \! \left( \frac{\pi}{2d} \right) \! \right\} \! \! \! \! \! \! \\
    \! \! \! \! \! \! g\left(\acute{z}_{q+N_t,i}\right) \! \! &= \! \! \cos^{2}\left\{ \! \left( \sinh(\dot{\gamma}) \acute{\underline{y}}_{pi} + \cosh(\dot{\gamma}) \acute{\underline{y}}_{q+N_t,i} \right) \! \! \left( \frac{\pi}{2d} \right) \! \right\} \! \! \! \! \! \!
  \end{aligned}
\end{equation}
and the $2^{nd}$ and $3^{rd}$ terms of \eqref{MoenCostHyper} are obtained by replacing ($\dot{\gamma}$) with ($-\dot{\gamma}$) and indices `$p$' and `$q+N_t$' with `$q$' and `$p+N_t$' in \eqref{Hyper_AMA_Terms2}, respectively. Now, the modified objective function is used to find the optimum $(\dot{\gamma}^\circ)$. Matrices $\Hfd_{p,q+N_t}$ and $\Hfd_{q,p+N_t}$ are then computed using the above explained method and applied successively on $\acute{\underline{\Ym}}$. The process is repeated until convergence.

\subsection{Approximate Solution}\label{sec:Approx_Sol_Hyper}
Again we will use here the Taylor series approximation of trigonometric angles given in \eqref{ApproxCosSin} and hyperbolic angles around zero up to $4^{\text{th}}$ order, which can be written as
\begin{equation}\label{ApproxHyper}
    \sinh(\gamma) \approx \gamma + \frac{\gamma^3}{6}, \qquad \cosh(\gamma) \approx 1 + \frac{\gamma^2}{2} + \frac{\gamma^4}{24}
\end{equation}
Let's consider the first term of \eqref{MoenCostHyper}, which is given in \eqref{Hyper_AMA_Terms} as
\begin{equation}\label{gZpHyper}
    g\left(\acute{z}_{pi}\right) = 1 - \sin^{2}\left\{ \left( \cosh(\gamma) \acute{\underline{y}}_{pi} + \sinh(\gamma) \acute{\underline{y}}_{qi} \right) \left( \frac{\pi}{2d} \right) \right\}
\end{equation}
Now, applying the hyperbolic angle approximation given in \eqref{ApproxHyper} to `$\cosh(\gamma)$' and `$\sinh(\gamma)$' in the argument of sine in \eqref{gZpHyper} and expanding the terms, we get
\begin{multline}\label{gZpApprHyper}
    g\left(\acute{z}_{pi}\right) \approx 1 - \sin^{2}\left\{ \left( 24 \acute{\underline{y}}_{pi} + 24 \acute{\underline{y}}_{qi} \gamma + 12 \acute{\underline{y}}_{pi} \gamma^2 \right. \right. \\ \left. \left. + 4 \acute{\underline{y}}_{qi} \gamma^3 + \acute{\underline{y}}_{pi} \gamma^4 \right) \left( \frac{\pi}{12d} \right) \right\}
\end{multline}
Finally, the approximation in \eqref{ApproxCosSin} is used leading to
\begin{equation}\label{gZpAppr2Hyper}
    \! ~ g\left(\acute{z}_{pi}\right) \! \! \approx \! \! \frac{1}{48 d^4} c_4^{pi} \gamma^4 \!+\! \frac{1}{12 d^3} c_3^{pi} \gamma^3 \!-\! \frac{1}{4 d^2} c_2^{pi} \gamma^2 \!-\! \frac{1}{2d} c_1^{pi} \gamma \!+\! \frac{1}{2} c_0^{pi} \! \! \!
\end{equation}
where
\begin{equation}\label{coeffHyper}
    \begin{aligned}
        c_4^{pi} &= \pi^4 \acute{\underline{y}}_{qi}^4 \cos \left(\frac{\pi \acute{\underline{y}}_{pi}}{d}\right) + 6 \pi^3 d \acute{\underline{y}}_{pi} \acute{\underline{y}}_{qi}^2 \sin \left(\frac{\pi  \acute{\underline{y}}_{pi}}{d}\right)\\
        & - 4 \pi^2 d^2 \acute{\underline{y}}_{qi}^2 \cos \left(\frac{\pi \acute{\underline{y}}_{pi}}{d}\right) - 3 \pi^2 d^2 \acute{\underline{y}}_{pi}^2 \cos \left(\frac{\pi \acute{\underline{y}}_{pi}}{d}\right)\\
        & - \pi d^3 \acute{\underline{y}}_{pi} \sin \left(\frac{\pi \acute{\underline{y}}_{pi}}{d}\right)\\
        c_3^{pi} &= \pi ^3 \acute{\underline{y}}_{qi}^3 \sin \left(\frac{\pi  \acute{\underline{y}}_{pi}}{d}\right) - \pi  d^2 \acute{\underline{y}}_{qi} \sin \left(\frac{\pi  \acute{\underline{y}}_{pi}}{d}\right)\\
        &~~ -3 \pi ^2 d \acute{\underline{y}}_{pi} \acute{\underline{y}}_{qi} \cos \left(\frac{\pi  \acute{\underline{y}}_{pi}}{d}\right)\\
        c_2^{pi} &= \pi ^2 \acute{\underline{y}}_{qi}^2 \cos \left(\frac{\pi \acute{\underline{y}}_{pi}}{d}\right) + \pi  d \acute{\underline{y}}_{pi} \sin \left(\frac{\pi  \acute{\underline{y}}_{pi}}{d}\right)\\
        c_1^{pi} & = \pi  \acute{\underline{y}}_{qi} \sin \left(\frac{\pi  \acute{\underline{y}}_{pi}}{d}\right), \quad c_0^{pi} = 1 + \cos \left(\frac{\pi  \acute{\underline{y}}_{pi}}{d}\right)
    \end{aligned}
\end{equation}
Similarly, the other terms $g\left(\acute{z}_{qi}\right)$, $g\left(\acute{z}_{p+N_t,i}\right)$ and $g\left(\acute{z}_{q+N_t,i}\right)$ of \eqref{MoenCostHyper} can be approximated as \eqref{gZpAppr2Hyper}, where the coefficients are obtained by replacing indices accordingly in \eqref{coeffHyper}.

Now, using \eqref{gZpAppr2Hyper} in \eqref{MoenCostHyper} leads to
\begin{equation}\label{MoenCostPolyApprHyper}
    \! ~ \Jc_{\text{AMA}} \! \! \approx \! \! \frac{1}{48 d^4} C_4 \gamma^4 \!+\! \frac{1}{12 d^3} C_3 \gamma^3 \!-\! \frac{1}{4 d^2} C_2 \gamma^2 \!-\! \frac{1}{2d} C_1 \gamma^1 \!+\! \frac{1}{2} C_0 \! \! \! \!
\end{equation}
where the coefficients in \eqref{MoenCostPolyApprHyper} are given by
\begin{equation}\label{Coef_sum_Hyper}
    C_l = \sum_{i=1}^{N_s} \left( c_l^{pi} + c_l^{qi} + c_l^{p+N_t,i} + c_l^{q+N_t,i} \right)
\end{equation}
where $l \in \{ 0,\dots,4 \}$. Taking the gradient with respect to $(\gamma)$ of AMA cost function in \eqref{MoenCostPolyApprHyper}, we get
\begin{equation}\label{HGAMA_Sol}
    \frac{\partial \Jc_{\text{AMA}}(\gamma)}{\partial \gamma} \approx \frac{1}{12 d^4} C_4 \gamma^3 + \frac{1}{4 d^3} C_3 \gamma^2 - \frac{1}{2 d^2} C_2 \gamma - \frac{1}{2d} C_1
\end{equation}
Out of the three possible real roots of \eqref{HGAMA_Sol}, the optimum $(\gamma^\circ)$ is selected such that $\Jc_{\text{AMA}}(\gamma)$ is minimum.

\begin{figure}[tb!]\centering
    \setlength\figureheight{3.5cm}
	\setlength\figurewidth{8.2cm}
    \input{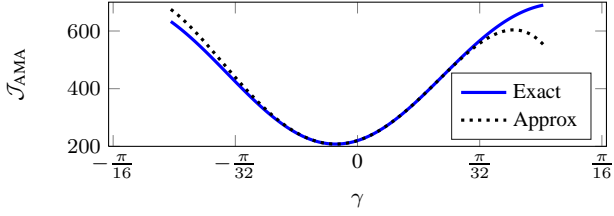}
    \caption{Comparison of exact and approximated H-AMA criterion after $n_0 = 5$ of G-MMA with $N_t = 3, N_r = 5, N_s = 300$, SNR$=30$dB and normalized 64-QAM constellation.}
    \label{CompCostHyper}
\end{figure}

To illustrate that the approximation in \eqref{MoenCostPolyApprHyper} is good enough, we have compared the original cost function and approximated one for a certain range of $(\gamma)$ around zero in Figure \ref{CompCostHyper}.

For the remaining matrices $\Hfd_{p,q+N_t}(\dot{\gamma})$ and $\Hfd_{q,p+N_t}(-\dot{\gamma})$, the AM cost function can be written as \eqref{MoenCostHyper} and all the terms have the same approximation as given in \eqref{gZpAppr2Hyper} with the replacement of $(\gamma)$ with $(\dot{\gamma})$ and indices accordingly. Also, for the $2^{\text{nd}}$ and $3^{\text{rd}}$ terms, the sign of coefficients $c_1$ and $c_3$ are opposite to the one shown in \eqref{gZpAppr2Hyper}. Now, using \eqref{gZpAppr2Hyper} the optimization problem in \eqref{MoenCostHyper} can be written as
\begin{equation}\label{MoenCostPolyApprHyper2}
    \! ~ \Jc_{\text{AMA}} \! \! \approx \! \! \frac{1}{48 d^4} C_4 \dot{\gamma}^4 \!+\! \frac{1}{12 d^3} C_3 \dot{\gamma}^3 \!-\! \frac{1}{4 d^2} C_2 \dot{\gamma}^2 \!+\! \frac{1}{2d} C_1 \dot{\gamma}^1 \!+\! \frac{1}{2} C_0 \! \! \! \!
\end{equation}
with
\begin{equation}\label{Coef_sum_Hyper2}
  \begin{aligned}
    \! C_l \!&=\! \sum_{i=1}^{N_s} \! \left( c_l^{pi} + c_l^{q+N_t,i} + c_l^{qi} + c_l^{p+N_t,i} \right) ~ l \in \{ 0,2,4 \} \\
    \! C_3 \!&=\! \sum_{i=1}^{N_s} \! \left( c_3^{pi} + c_3^{q+N_t,i} - c_3^{qi} - c_3^{p+N_t,i} \right) \\
    \! C_1 \!&=\! \sum_{i=1}^{N_s} \! \left( - c_1^{pi} - c_1^{q+N_t,i} + c_1^{qi} + c_1^{p+N_t,i} \right) \\ \!\!\!
  \end{aligned}
\end{equation}
The final solution is obtained by zeroing the gradient of \eqref{MoenCostPolyApprHyper2}. Once we obtain the solution $(\dot{\gamma}^\circ)$, matrices $\Hfd_{p,q+N_t}(\dot{\gamma}^\circ)$ and $\Hfd_{q,p+N_t}(-\dot{\gamma}^\circ)$  are computed using \eqref{HyperMatrix}. The separation matrix $\acute{\Vm}$ is then updated according to \eqref{HGV}.

In summary, matrix $\acute{\Vm}$ is initialized as identity matrix then after applying 5 sweeps of G-MMA, matrix $\acute{\Vm}$ is updated according to \eqref{HGV} by applying Givens and hyperbolic rotations successively on $\acute{\underline{\Ym}}$ using the above explained method, until convergence. The overall algorithm is summarized in Table \ref{HGAMAAlgo}.

\begin{table}[tb!]
\caption{Hyperbolic Givens AMA (HG-AMA) Algorithm}
\label{HGAMAAlgo}
\begin{boxedalgorithmic}
\STATE Initialization: $\acute{\Vm} = \Id_{2N_t}$
\STATE Subspace projection or approximate pre-whitening if $N_r>N_t$
\STATE 1. Construct real matrix $\acute{\underline{\Ym}}$ using (\ref{Vstruct})
\STATE 2. Hyperbolic $\&$ Givens Rotations:
\FOR{$n=1:N_{Sweeps}$}
    \IF{$n<=n_0$}
        \STATE a) Apply G-MMA as given in Table \ref{GMMAAlgo}
    \ELSE
        \FOR{$p=1:N_t-1$}
            \FOR{$q=p+1:N_t$}
                \STATE b) Find optimum ($\gamma^\circ$) using roots of \eqref{HGAMA_Sol} which gives minimum value of \eqref{MoenCostHyper}
                \STATE c) Compute $\Hfd_{p,q} \, \& \, \Hfd_{p+N_t,q+N_t}$ using \eqref{HyperMatrix} for same ($\gamma^\circ$)
                \STATE d) $\acute{\underline{\Ym}} = \Hfd_{p,q} \, \Hfd_{p+N_t,q+N_t} \acute{\underline{\Ym}}$
                \STATE e) $\acute{\Vm}=\Hfd_{p,q} \, \Hfd_{p+N_t,q+N_t} \acute{\Vm}$
                \STATE f) Apply Givens rotations using (b \TO e) of Table \ref{GAMAAlgo}
                \STATE \textbf{repeat} steps (b \TO f) for $(p,q+N_t) ~\&~ (q,p+N_t)$ using $(\dot{\theta}^\circ,\dot{\gamma}^\circ) ~\&~ (\dot{\theta}^\circ,-\dot{\gamma}^\circ)$, respectively
            \ENDFOR
        \ENDFOR
    \ENDIF
\ENDFOR
\end{boxedalgorithmic}
\end{table}

\section{Practical Considerations}
\label{sec:comments}

We provide here some insight into the proposed algorithms.

\subsection{Numerical Cost}\label{num_cost}
Taking into account the structure of the rotation matrices, the numerical cost of the proposed algorithms are compared with other CMA-like BSS algorithms in terms of the number of flops per sweep in Table \ref{NumCompBSS}.
\begin{table}[tb!]
\centering
\caption{Numerical complexity of different BSS algorithms}
\label{NumCompBSS}
\begin{tabular}{|c|c|}
\hline
\textbf{BSS Algorithm}  &  \textbf{Complexity Order}     \\ \hline
HG-AMA                  &  $140 N_s N_t^2 + \mathcal{O}(N_s N_t)$   \\ \hline
G-AMA                   &  $70 N_s N_t^2 + \mathcal{O}(N_s N_t)$   \\ \hline
HG-MMA                  &  $40 N_s N_t^2 + \mathcal{O}(N_s N_t)$   \\ \hline
G-MMA                   &  $20 N_s N_t^2 + \mathcal{O}(N_s N_t)$   \\ \hline
HG-CMA                  &  $30 N_s N_t^2 + \mathcal{O}(N_s N_t)$   \\ \hline
G-CMA                   &  $15 N_s N_t^2 + \mathcal{O}(N_s N_t)$   \\ \hline
ACMA                    &  $\mathcal{O}(N_s N_t^4)$           \\ \hline
\end{tabular}
\end{table}
As can be seen from Table \ref{NumCompBSS}, the proposed algorithms are much cheaper than ACMA and of the same cost order as G-CMA and HG-CMA. Moreover, the proposed algorithms have very fast convergence (typically less than 10 sweeps) as shown next in the simulation experiments. Also, HG-AMA is more expensive but has better performance than all the other algorithms as can be observed from the simulations results.

\subsection{Adaptive implementation}\label{adaptive_version}
The numerical cost of the designed batch algorithms increases linearly with the sample size $N_s$. Furthermore, in real life environments, systems are time varying and hence the separation matrix $\Wm$ has to be re-estimated or updated along the time axis. Thus, for slowly time varying systems, this update can be obtained by using adaptive estimation methods. Utilizing a sliding window technique as in \cite{GandHGCMA}, one can achieve such source separation in an adaptive manner with a numerical cost proportional to $\mathcal{O}(\acute{N_s}N_t^2)$ where $\acute{N_s}$ is the window size (instead of total sample size $N_s$).

\subsection{Complex implementation}\label{complex_imp}
As shown in section \ref{sec:motivation}, the real matrix representation has been introduced to overcome the difficulties encountered for the optimization of parameters of complex Givens and hyperbolic rotations. However, we can observe that the obtained results can be cast into complex matrix forms using the following straightforward relations:
\begin{equation}\label{Real2CompGH}
    \begin{aligned}
        \Gfd_{p,q}(\theta) \, \Gfd_{p+N_t,q+N_t}(\theta) \, \acute{\underline{\Ym}} &\iff \Gfd_{p,q}(\theta, 0) \underline{\Ym} \\
        \Hfd_{p,q}(\gamma) \, \Hfd_{p+N_t,q+N_t}(\gamma) \, \acute{\underline{\Ym}} &\iff \Hfd_{p,q}(\gamma, 0) \underline{\Ym} \\
        \Gfd_{p,q+N_t}(\dot{\theta}) \, \Gfd_{q,p+N_t}(\dot{\theta}) \, \acute{\underline{\Ym}} &\iff \Gfd_{p,q}(\theta, -\frac{\pi}{2}) \underline{\Ym} \\
        \Hfd_{p,q+N_t}(\acute{\gamma}) \, \Hfd_{q,p+N_t}(\acute{\gamma}) \, \acute{\underline{\Ym}} &\iff \Hfd_{p,q}(\gamma, -\frac{\pi}{2}) \underline{\Ym}
    \end{aligned}
\end{equation}
where all matrices on left side of (\ref{Real2CompGH}) are real and the right ones are complex. Somehow, we have replaced the two degrees of freedom of $\Gfd_{p,q}(\theta,\alpha)$ (resp. $\Hfd_{p,q}(\gamma,\beta)$) by the two free parameters $\theta$ and $\dot{\theta}$ (resp. $\gamma$ and $\acute{\gamma}$). This way we have avoided the non-linear optimization discussed in section \ref{sec:motivation}.

\subsection{Performance}\label{perf}
The main advantage of the proposed algorithms resides in their fast convergence in terms of the number of sweeps (typically less than 10 sweeps are needed for convergence) and also in terms of sample size (typically $N_s=\mathcal{O}(10N_t)$ is sufficient for the algorithm's convergence). Comparatively, the ACMA method requires $N_s=\mathcal{O}(10N_t^2)$ samples for its convergence and standard CMA-like methods need even more samples to converge to their steady state.

\section{Simulation Results}\label{sec:Sim_Res}
In order to evaluate the performance of the proposed algorithms, simulation results are presented in this section. Due to the lack of any batch BSS algorithm dealing with the MM criterion, we do the comparison with batch BSS algorithms dealing with the CM criterion such as ACMA, G-CMA and HG-CMA w.r.t. convergence rate, SER and SINR defined by
\begin{equation}\label{SINRformula}
    \text{SINR} = \frac{1}{N_t} \sum_{j=1}^{N_t}{ \text{SINR}_j }
\end{equation}
with
\begin{equation}\label{SINRjformula}
    \text{SINR}_j = \frac{|g_{jj} \sv_j|^2/N_s}{\sum_{l,l \neq j}{ |g_{jl} \sv_l|^2 }/N_s + \wv_j \Rm_n \wv_j^\herm}
\end{equation}
where $\text{SINR}_j$ is the signal to interference and noise ratio at the $j$th output with $g_{ij} = \wv_i \av_j$, where $\wv_i$ and $\av_j$ are the $i$th row vector and $j$th column vector of separation matrix $\Wm$ and mixing matrix $\Am$, respectively. $\Rm_n$ is the noise covariance matrix and $\sv_j$ is the ($1 \times N_s$) source vector at $j$th input.

We consider a MIMO system consisting of $5$ transmitters and $7$ receivers $(N_t=5, N_r=7)$ with the data model given in Section \ref{sec:Prob_Form}. Every uncoded data symbol transmitted by each source is drawn from $16$-QAM, $64$-QAM and $256$-QAM constellations. The resulting signals are then passed through matrix $\Am$, generated randomly at each Monte Carlo run with controlled conditioning and with i.i.d complex Gaussian variable entries of zero mean and unity variance. The noise variance is adjusted according to specified signal to noise ratio (SNR). The results are averaged over $1000$ Monte Carlo runs.

\subsection{Experiment 1: Exact vs. Approximate Solution of HG-MMA}\label{Exp1}
\begin{figure}[tb!]\centering
	\setlength\figureheight{5.5cm}
	\setlength\figurewidth{8.2cm}
	\footnotesize
\begin{tikzpicture}
\begin{axis}[
width=\figurewidth,
height=\figureheight,
xmin=0, xmax=30,
xlabel={SNR (dB)},
xlabel near ticks,
ymin=-10, ymax=25,
ylabel={Average SINR (dB)},
ylabel near ticks,
legend style={at={(0.98,0.02)},anchor=south east,legend cell align=left,align=left,draw=black}
]
\addplot [color=black,line width=1.2pt,mark size=3.0pt,only marks,mark=x,mark options={solid}]
  table[row sep=crcr]{%
0	-5.04812483421192\\
5	-0.136576095121275\\
10	6.60723503027659\\
15	11.8277737805011\\
20	15.8567759700992\\
25	18.933099894416\\
30	20.9171710903454\\
};
\addlegendentry{HG-MMA: exact};

\addplot [color=red,line width=1.2pt,mark size=4.5pt,only marks,mark=o,mark options={solid}]
  table[row sep=crcr]{%
0	-5.04707592308808\\
5	-0.151360596085128\\
10	6.59580145742189\\
15	11.7927333251728\\
20	15.8085864974587\\
25	18.8455449553863\\
30	20.7850813513008\\
};
\addlegendentry{HG-MMA: approx};

\addplot [color=black,solid,line width=1.2pt]
  table[row sep=crcr]{%
0	-5.04812483421192\\
5	-0.136576095121275\\
10	6.60723503027659\\
15	11.8277737805011\\
20	15.8567759700992\\
25	18.933099894416\\
30	20.9171710903454\\
};
\addlegendentry{16-QAM};

\addplot [color=black,line width=1.2pt,mark size=3.0pt,only marks,mark=x,mark options={solid},forget plot]
  table[row sep=crcr]{%
0	-5.09004774706618\\
5	-0.634770820533105\\
10	5.54478656780515\\
15	10.8671372035381\\
20	14.5881905005201\\
25	17.052763825706\\
30	18.4190946298482\\
};
\addplot [color=black,dashed,line width=1.2pt]
  table[row sep=crcr]{%
0	-5.09004774706618\\
5	-0.634770820533105\\
10	5.54478656780515\\
15	10.8671372035381\\
20	14.5881905005201\\
25	17.052763825706\\
30	18.4190946298482\\
};
\addlegendentry{64-QAM};

\addplot [color=red,dashed,line width=1.2pt,mark size=4.5pt,only marks,mark=o,mark options={solid},forget plot]
  table[row sep=crcr]{%
0	-5.08726510264454\\
5	-0.633863808308392\\
10	5.53295110432122\\
15	10.8676485063475\\
20	14.5785206723071\\
25	17.011005055571\\
30	18.3836920795738\\
};
\end{axis}
\end{tikzpicture}%
	\caption{Average SINR of exact and approximate solution of HG-MMA vs. SNR for $N_t=5, N_r=7, N_s=100$ and $N_{Sweeps}=10$ considering both $16$-QAM and $64$-QAM.}
	\label{fig:fig1}
\end{figure}
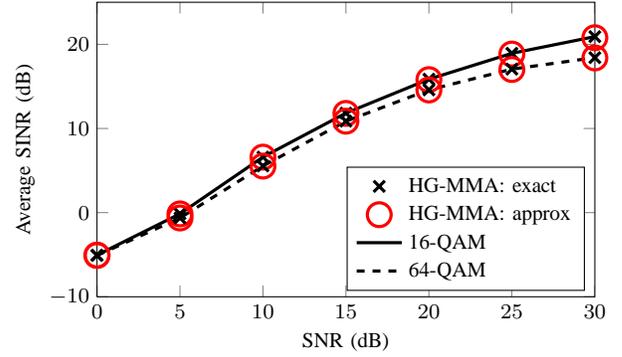

In Figure \ref{fig:fig1}, we compare the exact and approximate solution of HG-MMA in terms of SINR vs. SNR for $16$-QAM and $64$-QAM constellations. The number of sweeps $N_{Sweeps}$ and samples $N_s$ are set equal to $10$ and $100$, respectively. We notice that both the exact and approximate solutions have the same performance for the considered constellations. Therefore, in the following simulations for the HG-MMA, we will use the approximate solution, as it is cheaper and easier to implement.

\subsection{Experiment 2: Finding Optimum Number of Sweeps for G-MMA and HG-MMA}\label{Exp2}
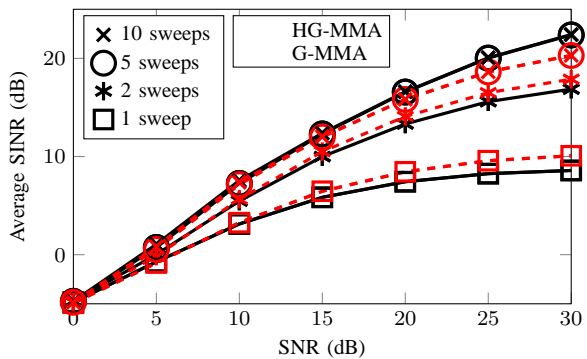
\begin{figure}[tb!]\centering
	\setlength\figureheight{5.5cm}
	\setlength\figurewidth{8.2cm}
	\footnotesize
\begin{tikzpicture}
\begin{axis}[
width=\figurewidth,
height=\figureheight,
xmin=0, xmax=30,
xlabel={SNR (dB)},
xlabel near ticks,
ymin=-5, ymax=25,
ylabel={Average SINR (dB)},
ylabel near ticks,
legend style={at={(0.02,0.98)},anchor=north west,legend cell align=left,align=left,draw=black}
]

\addplot [color=black,line width=1.2pt,mark size=3.5pt,only marks,mark=x,mark options={solid}]
  table[row sep=crcr]{%
0	-4.74825787094621\\
5	1.06276358869248\\
10	7.55478106496459\\
15	12.3779177125657\\
20	16.6005485557994\\
25	20.0611678346686\\
30	22.421248236041\\
};
\addlegendentry{10 sweeps};

\addplot [color=black,line width=1.2pt,mark size=4.5pt,only marks,mark=o,mark options={solid}]
  table[row sep=crcr]{%
0	-4.75158624251911\\
5	0.736219458217189\\
10	7.26183092997086\\
15	12.3006346697928\\
20	16.5677749549717\\
25	20.0243012428651\\
30	22.3836548077173\\
};
\addlegendentry{5 sweeps};

\addplot [color=black,line width=1.2pt,mark size=3.5pt,only marks,mark=asterisk,mark options={solid}]
  table[row sep=crcr]{%
0	-4.74534979734462\\
5	0.0310693659810351\\
10	5.46335880941832\\
15	9.98937439478051\\
20	13.3151539630932\\
25	15.5891223034895\\
30	16.8683051915125\\
};
\addlegendentry{2 sweeps};

\addplot [color=black,line width=1.2pt,mark size=3.5pt,only marks,mark=square,mark options={solid}]
  table[row sep=crcr]{%
0	-4.84383279638506\\
5	-0.722219638037617\\
10	3.1291941199726\\
15	5.8436420830115\\
20	7.43434745694425\\
25	8.25488231358467\\
30	8.58005008358981\\
};
\addlegendentry{1 sweep};

\addplot [color=black,solid,line width=1.2pt]
  table[row sep=crcr]{%
0	-4.84383279638506\\
5	-0.722219638037617\\
10	3.1291941199726\\
15	5.8436420830115\\
20	7.43434745694425\\
25	8.25488231358467\\
30	8.58005008358981\\
};\label{Lab_fig2_1}

\addplot [color=red,dashed,line width=1.2pt]
  table[row sep=crcr]{%
0	-4.94335537789675\\
5	-0.848933535175147\\
10	3.26178024336011\\
15	6.46391363470383\\
20	8.44509187159189\\
25	9.57542359861801\\
30	10.0763164639571\\
};\label{Lab_fig2_2}

\addplot [color=black,solid,line width=1.2pt,mark size=3.5pt,mark=square,mark options={solid},forget plot]
  table[row sep=crcr]{%
0	-4.84383279638506\\
5	-0.722219638037617\\
10	3.1291941199726\\
15	5.8436420830115\\
20	7.43434745694425\\
25	8.25488231358467\\
30	8.58005008358981\\
};
\addplot [color=red,dashed,line width=1.2pt,mark size=3.5pt,mark=square,mark options={solid},forget plot]
  table[row sep=crcr]{%
0	-4.94335537789675\\
5	-0.848933535175147\\
10	3.26178024336011\\
15	6.46391363470383\\
20	8.44509187159189\\
25	9.57542359861801\\
30	10.0763164639571\\
};
\addplot [color=black,solid,line width=1.2pt,mark size=3.5pt,mark=asterisk,mark options={solid},forget plot]
  table[row sep=crcr]{%
0	-4.74534979734462\\
5	0.0310693659810351\\
10	5.46335880941832\\
15	9.98937439478051\\
20	13.3151539630932\\
25	15.5891223034895\\
30	16.8683051915125\\
};
\addplot [color=red,dashed,line width=1.2pt,mark size=3.5pt,mark=asterisk,mark options={solid},forget plot]
  table[row sep=crcr]{%
0	-4.84915478799765\\
5	-0.146051553863098\\
10	5.64909078684069\\
15	10.4994968064966\\
20	14.0768738316498\\
25	16.5031663138086\\
30	17.8748087542869\\
};
\addplot [color=black,solid,line width=1.2pt,mark size=4.5pt,mark=o,mark options={solid},forget plot]
  table[row sep=crcr]{%
0	-4.75158624251911\\
5	0.736219458217189\\
10	7.26183092997086\\
15	12.3006346697928\\
20	16.5677749549717\\
25	20.0243012428651\\
30	22.3836548077173\\
};
\addplot [color=red,dashed,line width=1.2pt,mark size=4.5pt,mark=o,mark options={solid},forget plot]
  table[row sep=crcr]{%
0	-4.84701980767741\\
5	0.535337706190109\\
10	7.13753636146443\\
15	11.986791628436\\
20	15.8105576803846\\
25	18.6169254013072\\
30	20.2927105689689\\
};
\addplot [color=black,solid,line width=1.2pt,mark size=3.5pt,mark=x,mark options={solid},forget plot]
  table[row sep=crcr]{%
0	-4.74825787094621\\
5	1.06276358869248\\
10	7.55478106496459\\
15	12.3779177125657\\
20	16.6005485557994\\
25	20.0611678346686\\
30	22.421248236041\\
};
\addplot [color=red,dashed,line width=1.2pt,mark size=3.5pt,mark=x,mark options={solid},forget plot]
  table[row sep=crcr]{%
0	-4.84397815688327\\
5	0.812548312543926\\
10	7.37166811909711\\
15	12.0284529935127\\
20	15.8243900644383\\
25	18.6234114101617\\
30	20.3015348182571\\
};

\node [draw,fill=white,anchor=north west] at (rel axis cs: 0.32,0.98) {\shortstack[l]{
        \ref{Lab_fig2_1} ~~~~~ HG-MMA \\
        \ref{Lab_fig2_2} ~~~~~ G-MMA}};

\end{axis}
\end{tikzpicture}%
	\caption{Average SINR of HG-MMA and G-MMA vs. SNR for different $N_{Sweeps}$ considering $N_t=5, N_r=7, N_s=150$ and $16$-QAM constellation.}
	\label{fig:fig2}
\end{figure}

In Figure \ref{fig:fig2}, we examine the effect of the number of sweeps on the SINR of the G-MMA and HG-MMA for the case of $N_s = 150$ and $16$-QAM. We notice that the performance of proposed algorithms increases with the number of sweeps and remains almost unchanged after $5$ sweeps. So, in the following simulations we will fix the number of sweeps to $5$.

\subsection{Experiment 3: Exact vs. Approximate Solution of G-AMA and HG-AMA}\label{Exp3}
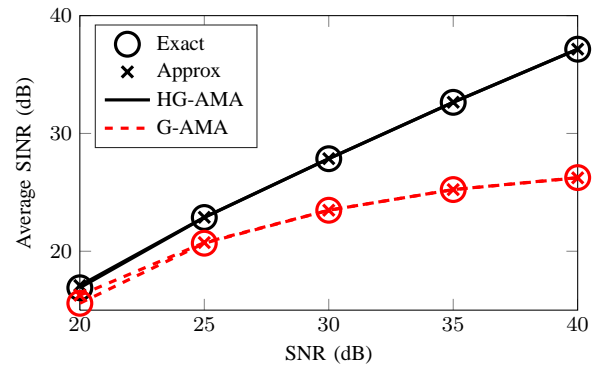
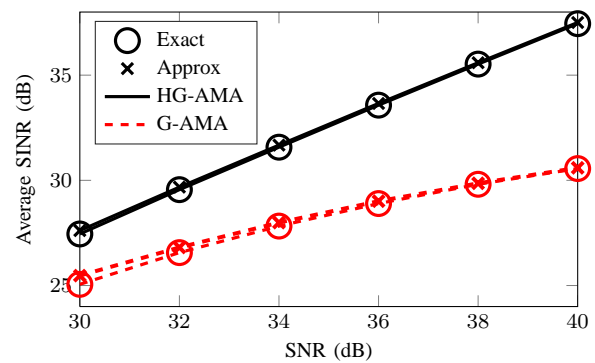
\begin{figure}[tb!]\centering
	\setlength\figureheight{5.5cm}
	\setlength\figurewidth{8.2cm}
    \subfloat[$64$-QAM, $N_s=200$]{\label{fig:fig3a}
	\footnotesize
\begin{tikzpicture}
\begin{axis}[%
width=\figurewidth,
height=\figureheight,
xmin=20, xmax=40,
xlabel={SNR (dB)},
xlabel near ticks,
ymin=15,ymax=40,
ylabel={Average SINR (dB)},
ylabel near ticks,
legend style={at={(0.03,0.97)},anchor=north west,legend cell align=left,align=left,draw=white!15!black}
]
\addplot [color=black,line width=1.2pt,mark size=4.5pt,only marks,mark=o,mark options={solid}]
  table[row sep=crcr]{%
20	16.8987323488092\\
25	22.8658335668714\\
30	27.8546840618042\\
35	32.6402554694388\\
40	37.1510983060758\\
};
\addlegendentry{Exact};

\addplot [color=black,line width=1.2pt,mark size=3.0pt,only marks,mark=x,mark options={solid}]
  table[row sep=crcr]{%
20	17.0885980417899\\
25	22.8813499708405\\
30	27.8601829229545\\
35	32.6489479545793\\
40	37.1543435772473\\
};
\addlegendentry{Approx};

\addplot [color=black,solid,line width=1.2pt]
  table[row sep=crcr]{%
20	16.8987323488092\\
25	22.8658335668714\\
30	27.8546840618042\\
35	32.6402554694388\\
40	37.1510983060758\\
};
\addlegendentry{HG-AMA};

\addplot [color=black,solid,line width=1.2pt,forget plot]
  table[row sep=crcr]{%
20	17.0885980417899\\
25	22.8813499708405\\
30	27.8601829229545\\
35	32.6489479545793\\
40	37.1543435772473\\
};
\addplot [color=red,dashed,line width=1.2pt]
  table[row sep=crcr]{%
20	15.5833803858153\\
25	20.6829233246604\\
30	23.4799829502695\\
35	25.2360590436026\\
40	26.2508917753481\\
};
\addlegendentry{G-AMA};

\addplot [color=red,line width=1.2pt,mark size=4.5pt,only marks,mark=o,mark options={solid},forget plot]
  table[row sep=crcr]{%
20	15.5833803858153\\
25	20.6829233246604\\
30	23.4799829502695\\
35	25.2360590436026\\
40	26.2508917753481\\
};
\addplot [color=red,dashed,line width=1.2pt,mark size=3.0pt,mark=x,mark options={solid},forget plot]
  table[row sep=crcr]{%
20	16.2665173029241\\
25	20.7342125860787\\
30	23.4827198426392\\
35	25.2317523279779\\
40	26.2314720428446\\
};
\end{axis}
\end{tikzpicture}%
} \vfill \medskip
    \subfloat[$256$-QAM, $N_s=500$]{\label{fig:fig3b}
	\footnotesize
\begin{tikzpicture}
\begin{axis}[%
width=\figurewidth,
height=\figureheight,
xmin=30, xmax=40,
xlabel={SNR (dB)},
xlabel near ticks,
ymin=24, ymax=38,
ylabel={Average SINR (dB)},
ylabel near ticks,
legend style={at={(0.03,0.97)},anchor=north west,legend cell align=left,align=left,draw=white!15!black}
]
\addplot [color=black,line width=1.2pt,mark size=4.5pt,only marks,mark=o,mark options={solid}]
  table[row sep=crcr]{%
30	27.4505734483717\\
32	29.5570052346335\\
34	31.575642910015\\
36	33.5740257927164\\
38	35.525074981597\\
40	37.4469435533099\\
};
\addlegendentry{Exact};

\addplot [color=black,line width=1.2pt,mark size=3.0pt,only marks,mark=x,mark options={solid}]
  table[row sep=crcr]{%
30	27.6113941978216\\
32	29.676450354935\\
34	31.6742809280029\\
36	33.6543789320081\\
38	35.5912328362568\\
40	37.5017889537658\\
};
\addlegendentry{Approx};

\addplot [color=black,solid,line width=1.2pt]
  table[row sep=crcr]{%
30	27.4505734483717\\
32	29.5570052346335\\
34	31.575642910015\\
36	33.5740257927164\\
38	35.525074981597\\
40	37.4469435533099\\
};
\addlegendentry{HG-AMA};

\addplot [color=black,solid,line width=1.2pt,forget plot]
  table[row sep=crcr]{%
30	27.6113941978216\\
32	29.676450354935\\
34	31.6742809280029\\
36	33.6543789320081\\
38	35.5912328362568\\
40	37.5017889537658\\
};
\addplot [color=red,dashed,line width=1.2pt]
  table[row sep=crcr]{%
30	25.0563332015128\\
32	26.5638016940347\\
34	27.8252237018173\\
36	28.8960387586405\\
38	29.8128831476574\\
40	30.5589579539522\\
};
\addlegendentry{G-AMA};

\addplot [color=red,line width=1.2pt,mark size=4.5pt,only marks,mark=o,mark options={solid},forget plot]
  table[row sep=crcr]{%
30	25.0563332015128\\
32	26.5638016940347\\
34	27.8252237018173\\
36	28.8960387586405\\
38	29.8128831476574\\
40	30.5589579539522\\
};
\addplot [color=red,dashed,line width=1.5pt,mark size=3.0pt,mark=x,mark options={solid},forget plot]
  table[row sep=crcr]{%
30	25.4711484659502\\
32	26.8096267340696\\
34	27.9896694814448\\
36	29.0092455570659\\
38	29.8588177830722\\
40	30.5944325287646\\
};
\end{axis}
\end{tikzpicture}%
}
	\caption{Average SINR of exact and approximate solution of HG-AMA and G-AMA vs. SNR for $N_t=5$, $N_r=7$, $N_{Sweeps}=10$.}
	\label{fig:fig3}
\end{figure}

Now, we compare the performance of exact and approximate solutions presented for G-AMA and HG-AMA in terms of SINR vs. SNR. Figure \ref{fig:fig3a} and \ref{fig:fig3b} shows the plots for $N_s=200$, 64-QAM and $N_s=500$, 256-QAM constellations, respectively. The number of sweeps $N_{Sweeps}$ is fixed at 10, where we used 5 sweeps of G-MMA followed by 5 sweeps of AMAs. From Figure \ref{fig:fig3}, we notice that both the exact and approximate solutions have the same performance. Therefore, in the following simulations for the G-AMA and HG-AMA, we will use the approximate solution, as it is cheaper and easier to implement.

\subsection{Experiment 4: Finding Optimum Number of Sweeps for G-AMA and HG-AMA}\label{Exp4}
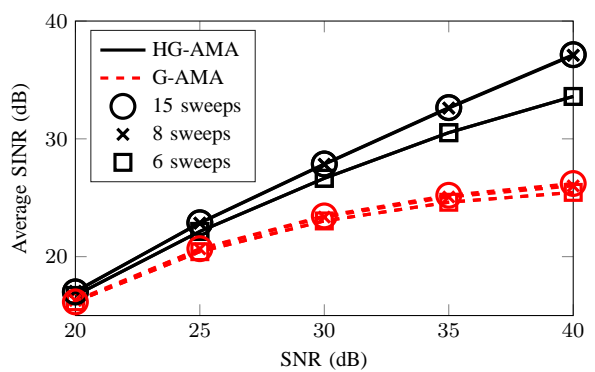
\begin{figure}[tb!]\centering
	\setlength\figureheight{5.5cm}
	\setlength\figurewidth{8.2cm}
	\footnotesize
\begin{tikzpicture}
\begin{axis}[%
width=\figurewidth,
height=\figureheight,
xmin=20, xmax=40,
xlabel={SNR (dB)},
xlabel near ticks,
ymin=15, ymax=40,
ylabel={Average SINR (dB)},
ylabel near ticks,
legend style={at={(0.03,0.97)},anchor=north west,legend cell align=left,align=left,draw=white!15!black}
]
\addplot [color=black,solid,line width=1.2pt]
  table[row sep=crcr]{%
20	16.7052920369187\\
25	22.141241377391\\
30	26.6491929835399\\
35	30.5252617973651\\
40	33.5983158878937\\
};
\addlegendentry{HG-AMA};

\addplot [color=red,dashed,line width=1.2pt]
  table[row sep=crcr]{%
20	16.1734894973276\\
25	20.4225824582787\\
30	23.0311598813525\\
35	24.6182224273064\\
40	25.4566710969309\\
};
\addlegendentry{G-AMA};

\addplot [color=black,line width=1.2pt,mark size=4.5pt,only marks,mark=o,mark options={solid}]
  table[row sep=crcr]{%
20	17.0085197004399\\
25	22.8760022738408\\
30	27.867509807042\\
35	32.6513293468938\\
40	37.1435056889308\\
};
\addlegendentry{15 sweeps};

\addplot [color=black,line width=1.2pt,mark size=3.0pt,only marks,mark=x,mark options={solid}]
  table[row sep=crcr]{%
20	17.0534849861698\\
25	22.845236755743\\
30	27.8378614887546\\
35	32.608303718422\\
40	37.0916549921069\\
};
\addlegendentry{8 sweeps};

\addplot [color=black,line width=1.2pt,mark size=3.0pt,only marks,mark=square,mark options={solid}]
  table[row sep=crcr]{%
20	16.7052920369187\\
25	22.141241377391\\
30	26.6491929835399\\
35	30.5252617973651\\
40	33.5983158878937\\
};
\addlegendentry{6 sweeps};

\addplot [color=black,solid,line width=1.2pt,mark size=3.0pt,mark=square,mark options={solid},forget plot]
  table[row sep=crcr]{%
20	16.7052920369187\\
25	22.141241377391\\
30	26.6491929835399\\
35	30.5252617973651\\
40	33.5983158878937\\
};
\addplot [color=red,dashed,line width=1.2pt,mark size=3.0pt,mark=square,mark options={solid},forget plot]
  table[row sep=crcr]{%
20	16.1734894973276\\
25	20.4225824582787\\
30	23.0311598813525\\
35	24.6182224273064\\
40	25.4566710969309\\
};
\addplot [color=black,solid,line width=1.2pt,mark size=3.0pt,mark=x,mark options={solid},forget plot]
  table[row sep=crcr]{%
20	17.0534849861698\\
25	22.845236755743\\
30	27.8378614887546\\
35	32.608303718422\\
40	37.0916549921069\\
};
\addplot [color=red,dashed,line width=1.2pt,mark size=3.0pt,mark=x,mark options={solid},forget plot]
  table[row sep=crcr]{%
20	16.2803893526006\\
25	20.6575012339774\\
30	23.3589603930059\\
35	25.0649529387665\\
40	26.0070453484539\\
};
\addplot [color=black,solid,line width=1.2pt,mark size=4.5pt,mark=o,mark options={solid},forget plot]
  table[row sep=crcr]{%
20	17.0085197004399\\
25	22.8760022738408\\
30	27.867509807042\\
35	32.6513293468938\\
40	37.1435056889308\\
};
\addplot [color=red,dashed,line width=1.2pt,mark size=4.5pt,mark=o,mark options={solid},forget plot]
  table[row sep=crcr]{%
20	16.1584292776625\\
25	20.6760994943642\\
30	23.4582326769918\\
35	25.2087243415592\\
40	26.2098244052543\\
};
\end{axis}
\end{tikzpicture}%
	\caption{Average SINR of HG-AMA and G-AMA vs. SNR for different $N_{Sweeps}$ considering $N_t=5, N_r=7, N_s=200$ and $64$-QAM constellation.}
	\label{fig:fig4}
\end{figure}

In Figure \ref{fig:fig4}, we examine the effect of the number of sweeps $N_{Sweeps}$ on the SINR of the G-AMA and HG-AMA for the case of $N_s=200$ and 64-QAM. We notice that the performance of proposed algorithms increases with the number of sweeps and remains almost unchanged after $8$ sweeps (5 G-MMA + 3 AMA sweeps). So, in the following simulations we will fix the number of sweeps to $8$.

\subsection{Experiment 5: Comparison of Rate of Convergence}\label{Exp5}
In Figure \ref{fig:fig5}, we have compared the convergence rate of the the proposed and benchmarked algorithms. The SNR is fixed at 30 dB and $N_s$ is selected as 200 and 500 for 64-QAM and 256-QAM, respectively. It can be noticed that G-AMA and HG-AMA converge in 8 sweeps, while all other algorithms converge in 5 sweeps. Even though the proposed algorithms G-AMA and HG-AMA require 3 extra sweeps, the performance is much better than all the other algorithms. Moreover, the performance of HG-MMA and G-MMA is better than the HG-CMA and G-CMA.
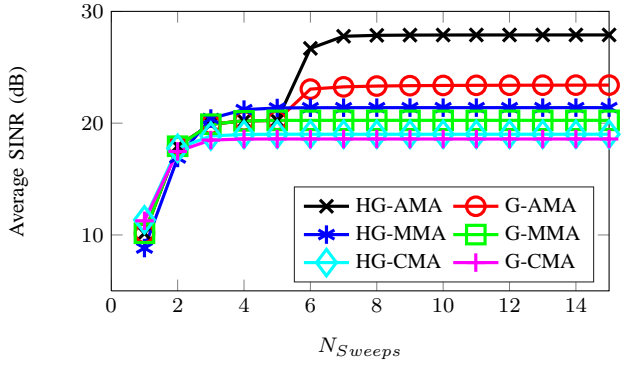
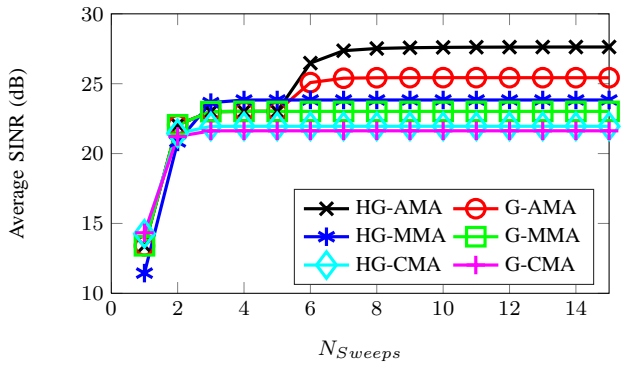
\begin{figure}[tb!]\centering
	\setlength\figureheight{5.3cm}
	\setlength\figurewidth{8.2cm}
    \subfloat[$64$-QAM, $N_s=200$]{\label{fig:fig5}
	\footnotesize
\definecolor{mycolor1}{rgb}{0.00000,1.00000,1.00000}%
\definecolor{mycolor2}{rgb}{1.00000,0.00000,1.00000}%
\begin{tikzpicture}
\begin{axis}[
width=\figurewidth,
height=\figureheight,
xmin=0,xmax=15,
xlabel={$N_{Sweeps}$},
ymin=5,ymax=30,
ylabel={Average SINR (dB)},
legend columns=2,
legend style={at={(0.97,0.03)},anchor=south east,legend cell align=left,align=left,draw=white!15!black}
]
\addplot [color=black,solid,line width=1.2pt,mark size=3.5pt,mark=x,mark options={solid}]
  table[row sep=crcr]{%
1	10.1520427897075\\
2	17.9376590916744\\
3	19.9067267330603\\
4	20.208005926146\\
5	20.2486980780789\\
6	26.6915459726055\\
7	27.773497591412\\
8	27.8572971103682\\
9	27.8782430388241\\
10	27.8863720407325\\
11	27.8903016780291\\
12	27.891799813924\\
13	27.8927193956069\\
14	27.8928494422787\\
15	27.8929090803071\\
};
\addlegendentry{HG-AMA};

\addplot [color=red,solid,line width=1.2pt,mark size=3.5pt,mark=o,mark options={solid}]
  table[row sep=crcr]{%
1	10.1520427897075\\
2	17.9376590916744\\
3	19.9067267330603\\
4	20.208005926146\\
5	20.2486980780789\\
6	23.0477322579543\\
7	23.2483780181151\\
8	23.320772220757\\
9	23.3541974686331\\
10	23.37394428121\\
11	23.3856926683521\\
12	23.3931754760027\\
13	23.3982170409278\\
14	23.4022217992979\\
15	23.4045150026348\\
};
\addlegendentry{G-AMA};

\addplot [color=blue,solid,line width=1.2pt,mark size=3.5pt,mark=asterisk,mark options={solid}]
  table[row sep=crcr]{%
1	8.82936792987682\\
2	16.8907718927267\\
3	20.4133294163666\\
4	21.2121208446846\\
5	21.3568147487337\\
6	21.3816389467058\\
7	21.3863649677769\\
8	21.3875798410728\\
9	21.3890594569225\\
10	21.3889053791944\\
11	21.3888266340025\\
12	21.3888228646548\\
13	21.3888247118881\\
14	21.38882536723\\
15	21.3888255146569\\
};
\addlegendentry{HG-MMA};

\addplot [color=green,solid,line width=1.2pt,mark size=3.5pt,mark=square,mark options={solid}]
  table[row sep=crcr]{%
1	10.1520427897075\\
2	17.9376590916744\\
3	19.9067267330603\\
4	20.208005926146\\
5	20.2486980780789\\
6	20.2520964313131\\
7	20.2536598377419\\
8	20.2536884555644\\
9	20.2536641339858\\
10	20.253658916856\\
11	20.2536579890577\\
12	20.2536578110278\\
13	20.2536577728756\\
14	20.2536577639615\\
15	20.2536577617167\\
};
\addlegendentry{G-MMA};

\addplot [color=mycolor1,solid,line width=1.2pt,mark size=5.0pt,mark=diamond,mark options={solid}]
  table[row sep=crcr]{%
1	11.3530562269817\\
2	17.7418595139394\\
3	18.8548006987872\\
4	18.9922377419855\\
5	19.0112352249992\\
6	19.0103132026227\\
7	19.008403032945\\
8	19.0081124270251\\
9	19.0080748332404\\
10	19.0084788025843\\
11	19.0089891048557\\
12	19.0091828666492\\
13	19.009375522863\\
14	19.0100048651966\\
15	19.0104248629411\\
};
\addlegendentry{HG-CMA};

\addplot [color=mycolor2,solid,line width=1.2pt,mark size=3.5pt,mark=+,mark options={solid}]
  table[row sep=crcr]{%
1	11.2657549902539\\
2	17.4737035148859\\
3	18.4922224093548\\
4	18.5804671631769\\
5	18.5856767354119\\
6	18.5847492652349\\
7	18.5844972032461\\
8	18.5844451058145\\
9	18.5844363936597\\
10	18.5844358763637\\
11	18.5844364032086\\
12	18.5844367987921\\
13	18.5844370050375\\
14	18.5844371006655\\
15	18.5844371427787\\
};
\addlegendentry{G-CMA};

\end{axis}
\end{tikzpicture}%
} \vfill \medskip
    \subfloat[$256$-QAM, $N_s=500$]{\label{fig:fig5b}
	\footnotesize
\definecolor{mycolor1}{rgb}{0.00000,1.00000,1.00000}%
\definecolor{mycolor2}{rgb}{1.00000,0.00000,1.00000}%
\begin{tikzpicture}
\begin{axis}[
width=\figurewidth,
height=\figureheight,
xmin=0,xmax=15,
xlabel={$N_{Sweeps}$},
ymin=10,ymax=30,
ylabel={Average SINR (dB)},
legend columns=2,
legend style={at={(0.97,0.03)},anchor=south east,legend cell align=left,align=left,draw=white!15!black}
]

\addplot [color=black,solid,line width=1.2pt,mark size=3.5pt,mark=x,mark options={solid}]
  table[row sep=crcr]{%
1	13.4003448007442\\
2	22.058562497002\\
3	23.0001097866369\\
4	23.0173463689262\\
5	23.0174609790316\\
6	26.4744231768698\\
7	27.361213933671\\
8	27.520619751589\\
9	27.5742997446682\\
10	27.5993902003901\\
11	27.6125959495188\\
12	27.619640254654\\
13	27.6238044158516\\
14	27.6259257028435\\
15	27.6269796001124\\
};\label{Lab_Fig5b_1}
\addlegendentry{HG-AMA};

\addplot [color=red,solid,line width=1.2pt,mark size=3.5pt,mark=o,mark options={solid}]
  table[row sep=crcr]{%
1	13.4003448007442\\
2	22.058562497002\\
3	23.0001097866369\\
4	23.0173463689262\\
5	23.0174609790316\\
6	25.0797297680736\\
7	25.3912710832479\\
8	25.4248492839781\\
9	25.4313698823047\\
10	25.4326362466623\\
11	25.431628747857\\
12	25.4307207751538\\
13	25.4293430194045\\
14	25.4276785556623\\
15	25.425233313734\\
};\label{Lab_Fig5b_2}
\addlegendentry{G-AMA};

\addplot [color=blue,solid,line width=1.2pt,mark size=3.5pt,mark=asterisk,mark options={solid}]
  table[row sep=crcr]{%
1	11.4452478883315\\
2	20.8340731702732\\
3	23.6456383776732\\
4	23.8316334177065\\
5	23.8348319208556\\
6	23.8347142777963\\
7	23.8346884070621\\
8	23.8346847327639\\
9	23.8346842983768\\
10	23.8346842478095\\
11	23.8346842409777\\
12	23.834684239798\\
13	23.8346842395452\\
14	23.8346842394857\\
15	23.8346842394715\\
};
\addlegendentry{HG-MMA};

\addplot [color=green,solid,line width=1.2pt,mark size=3.5pt,mark=square,mark options={solid}]
  table[row sep=crcr]{%
1	13.4003448007442\\
2	22.058562497002\\
3	23.0001097866369\\
4	23.0173463689262\\
5	23.0174609790316\\
6	23.0174466608505\\
7	23.0174441852417\\
8	23.0174439391132\\
9	23.017443916739\\
10	23.0174439147069\\
11	23.0174439145176\\
12	23.0174439144991\\
13	23.0174439144972\\
14	23.017443914497\\
15	23.017443914497\\
};
\addlegendentry{G-MMA};

\addplot [color=mycolor1,solid,line width=1.2pt,mark size=5.0pt,mark=diamond,mark options={solid}]
  table[row sep=crcr]{%
1	14.2603980696927\\
2	21.4474390656811\\
3	21.9445096796505\\
4	21.9517064761028\\
5	21.950547315041\\
6	21.9508204539221\\
7	21.9517927123212\\
8	21.9504122825023\\
9	21.9517565024782\\
10	21.9518378930175\\
11	21.9517801560212\\
12	21.9517750548071\\
13	21.9517746523557\\
14	21.9517746262375\\
15	21.9504789641388\\
};
\addlegendentry{HG-CMA};

\addplot [color=mycolor2,solid,line width=1.2pt,mark size=3.5pt,mark=+,mark options={solid}]
  table[row sep=crcr]{%
1	14.3306199926101\\
2	21.2205554916049\\
3	21.6254899301484\\
4	21.6285703238795\\
5	21.6284516935021\\
6	21.628431625055\\
7	21.6284295312543\\
8	21.628429388301\\
9	21.6284293844869\\
10	21.628429385484\\
11	21.628429385752\\
12	21.6284293857997\\
13	21.6284293858073\\
14	21.6284293858084\\
15	21.6284293858086\\
};
\addlegendentry{G-CMA};

\end{axis}
\end{tikzpicture}%
}
	\caption{Average SINR of iterative batch BSS algorithms vs. $N_{Sweeps}$ for $N_t=5$, $N_r=7$ and SNR $=30$dB.}
	\label{fig:fig5}
\end{figure}

\subsection{Experiment 6: Effect of the Number of Samples}\label{Exp6}
\begin{figure}[tb!]\centering
	\setlength\figureheight{5.3cm}
	\setlength\figurewidth{8.2cm}
    \subfloat[$64$-QAM]{\label{fig:fig6a}
	\footnotesize
\definecolor{mycolor1}{rgb}{0.00000,1.00000,1.00000}%
\definecolor{mycolor2}{rgb}{1.00000,0.00000,1.00000}%
\begin{tikzpicture}
\begin{axis}[
width=\figurewidth,
height=\figureheight,
xmin=0,xmax=500,
xlabel={Number of Samples $N_s$},
ymin=0,ymax=30,
ylabel={Average SINR (dB)},
legend columns=2,
legend style={at={(0.97,0.03)},anchor=south east,legend cell align=left,align=left,draw=white!15!black}
]
\addplot [color=black,solid,line width=1.2pt,mark size=3.5pt,mark=x,mark options={solid}]
  table[row sep=crcr]{%
50	14.3207628512572\\
};
\addlegendentry{HG-AMA};

\addplot [color=red,solid,line width=1.2pt,mark size=3.5pt,mark=o,mark options={solid}]
  table[row sep=crcr]{%
50	13.0430987286392\\
};
\addlegendentry{G-AMA};

\addplot [color=blue,solid,line width=1.2pt,mark size=3.5pt,mark=asterisk,mark options={solid}]
  table[row sep=crcr]{%
50	12.2211412114258\\
};
\addlegendentry{HG-MMA};

\addplot [color=green,solid,line width=1.2pt,mark size=3.5pt,mark=square,mark options={solid}]
  table[row sep=crcr]{%
50	11.644872826999\\
};
\addlegendentry{G-MMA};

\addplot [color=mycolor1,solid,line width=1.2pt,mark size=3.5pt,mark=diamond,mark options={solid}]
  table[row sep=crcr]{%
50	10.2520944587892\\
};
\addlegendentry{HG-CMA};

\addplot [color=mycolor2,solid,line width=1.2pt,mark size=3.5pt,mark=+,mark options={solid}]
  table[row sep=crcr]{%
50	9.99071673752469\\
};
\addlegendentry{G-CMA};

\addplot [color=blue,solid,line width=1.2pt,mark size=3.5pt,mark=triangle,mark options={solid}]
  table[row sep=crcr]{%
50	4.72002537059385\\
};
\addlegendentry{ACMA};

\addplot [color=black,solid,line width=1.2pt,mark size=3.5pt,mark=x,mark options={solid},forget plot]
  table[row sep=crcr]{%
50	14.3207628512572\\
100	25.4141504639445\\
150	27.4109661615689\\
200	27.9078728230933\\
250	28.0109814548527\\
300	28.0707157250173\\
350	28.0766620001794\\
400	28.0444908515675\\
450	28.0444627304379\\
500	28.0731885855082\\
};
\addplot [color=red,solid,line width=1.2pt,mark size=3.5pt,mark=o,mark options={solid},forget plot]
  table[row sep=crcr]{%
50	13.0430987286392\\
100	21.8957561718185\\
150	23.0628025017308\\
200	23.4545110173501\\
250	23.7386662408517\\
300	24.0295571470245\\
350	24.1878130684926\\
400	24.3567368743929\\
450	24.6135461116092\\
500	24.8120186598656\\
};
\addplot [color=blue,solid,line width=1.2pt,mark size=3.5pt,mark=asterisk,mark options={solid},forget plot]
  table[row sep=crcr]{%
50	12.2211412114258\\
100	18.4671139732395\\
150	20.3503099549975\\
200	21.407055830169\\
250	22.2266341692221\\
300	22.8158614478655\\
350	23.201920596207\\
400	23.5307195226711\\
450	23.8920782267756\\
500	24.1537561001823\\
};
\addplot [color=green,solid,line width=1.2pt,mark size=3.5pt,mark=square,mark options={solid},forget plot]
  table[row sep=crcr]{%
50	11.644872826999\\
100	17.3641239470233\\
150	19.1756555613317\\
200	20.2970496468668\\
250	21.111415247324\\
300	21.772026661036\\
350	22.1665861526861\\
400	22.5230710503935\\
450	22.9277657512404\\
500	23.2030144261572\\
};
\addplot [color=mycolor1,solid,line width=1.2pt,mark size=5.0pt,mark=diamond,mark options={solid},forget plot]
  table[row sep=crcr]{%
50	10.2520944587892\\
100	15.7381211761019\\
150	17.7643577745692\\
200	18.996566239972\\
250	19.8896219746837\\
300	20.6064807467577\\
350	21.089824494214\\
400	21.5621001822292\\
450	21.9195616612735\\
500	22.2969356442006\\
};
\addplot [color=mycolor2,solid,line width=1.2pt,mark size=3.5pt,mark=+,mark options={solid},forget plot]
  table[row sep=crcr]{%
50	9.99071673752469\\
100	15.3775469182465\\
150	17.3413929959295\\
200	18.5432437086008\\
250	19.4677654961325\\
300	20.1979655741266\\
350	20.6777843327304\\
400	21.1462345296293\\
450	21.5280270329009\\
500	21.8905882089293\\
};
\addplot [color=blue,solid,line width=1.2pt,mark size=3.5pt,mark=triangle,mark options={solid},forget plot]
  table[row sep=crcr]{%
50	4.72002537059385\\
100	11.7503745967967\\
150	15.0313568937324\\
200	16.8222312614085\\
250	17.9189249566044\\
300	18.7835740183996\\
350	19.4173179986339\\
400	20.006251041078\\
450	20.4230167097552\\
500	20.8935773110458\\
};
\end{axis}
\end{tikzpicture}%
} \vfill \medskip
    \subfloat[$256$-QAM]{\label{fig:fig6b}
	\footnotesize
\definecolor{mycolor1}{rgb}{0.00000,1.00000,1.00000}%
\definecolor{mycolor2}{rgb}{1.00000,0.00000,1.00000}%
\begin{tikzpicture}
\begin{axis}[
width=\figurewidth,
height=\figureheight,
xmin=0,xmax=1000,
xlabel={Number of Samples $N_s$},
ymin=10,ymax=30,
ylabel={Average SINR (dB)},
legend columns=2,
legend style={at={(0.97,0.03)},anchor=south east,legend cell align=left,align=left,draw=white!15!black}
]
\addplot [color=black,solid,line width=1.2pt,mark size=3.5pt,mark=x,mark options={solid}]
  table[row sep=crcr]{%
100	15.5973355448264\\
150	19.4964702956361\\
200	22.4565076260284\\
250	24.0370408872408\\
300	25.4843520936508\\
350	26.3253302433592\\
400	26.9048675977736\\
500	27.5634184848136\\
600	27.8291732900817\\
700	27.9315132438348\\
800	28.0441189622836\\
900	28.0633542642836\\
1000	28.0271289745045\\
};
\addlegendentry{HG-AMA};

\addplot [color=red,solid,line width=1.2pt,mark size=3.5pt,mark=o,mark options={solid}]
  table[row sep=crcr]{%
100	15.4039368621204\\
150	18.5297716490338\\
200	21.3380801497994\\
250	22.7815044033676\\
300	23.8752855240918\\
350	24.376788692389\\
400	24.9179887317476\\
500	25.4335745862901\\
600	25.7245375031377\\
700	25.9593682084062\\
800	26.1318891589774\\
900	26.2814148910086\\
1000	26.3652696644889\\
};
\addlegendentry{G-AMA};

\addplot [color=blue,solid,line width=1.2pt,mark size=3.5pt,mark=asterisk,mark options={solid}]
  table[row sep=crcr]{%
100	17.7916977805704\\
150	19.8608778974366\\
200	21.0249559147596\\
250	21.7559574542288\\
300	22.3545920070328\\
350	22.8416557789248\\
400	23.2760757298344\\
500	23.8347148596803\\
600	24.2980321799631\\
700	24.6161507501464\\
800	24.9533201130447\\
900	25.1630603988684\\
1000	25.3586298461287\\
};
\addlegendentry{HG-MMA};

\addplot [color=green,solid,line width=1.2pt,mark size=3.5pt,mark=square,mark options={solid}]
  table[row sep=crcr]{%
100	16.9933616417956\\
150	18.829854919973\\
200	20.0049290939747\\
250	20.8068498259436\\
300	21.4579070648755\\
350	21.937873990206\\
400	22.3942196965968\\
500	23.019439139419\\
600	23.5058198518082\\
700	23.8861154222108\\
800	24.243558075957\\
900	24.4890521876358\\
1000	24.7238814505251\\
};
\addlegendentry{G-MMA};

\addplot [color=mycolor1,solid,line width=1.2pt,mark size=5.0pt,mark=diamond,mark options={solid}]
  table[row sep=crcr]{%
100	15.3661863790771\\
150	17.3863094297735\\
200	18.7656179063209\\
250	19.5250353017467\\
300	20.2101274619389\\
350	20.7411178485037\\
400	21.2401349384997\\
500	21.9693398564339\\
600	22.5138835949228\\
700	22.9346481644165\\
800	23.3773028332899\\
900	23.6369325123834\\
1000	23.8710143492528\\
};
\addlegendentry{HG-CMA};

\addplot [color=mycolor2,solid,line width=1.2pt,mark size=3.5pt,mark=+,mark options={solid}]
  table[row sep=crcr]{%
100	14.9795722160899\\
150	17.0661623555867\\
200	18.362144385349\\
250	19.1469987174595\\
300	19.8527868005179\\
350	20.411813762838\\
400	20.8837070194614\\
500	21.6353262801596\\
600	22.1944370201121\\
700	22.6405026670795\\
800	23.0729570761133\\
900	23.348839649741\\
1000	23.6080944899314\\
};
\addlegendentry{G-CMA};

\addplot [color=blue,solid,line width=1.2pt,mark size=3.5pt,mark=triangle,mark options={solid}]
  table[row sep=crcr]{%
100	11.2196252670982\\
150	14.5115977398403\\
200	16.453050145095\\
250	17.4309855402324\\
300	18.330770296425\\
350	19.0239276943293\\
400	19.6069709949366\\
500	20.4674941042716\\
600	21.1061264074361\\
700	21.6427608184464\\
800	22.1454092605906\\
900	22.4803196015581\\
1000	22.7550315967089\\
};
\addlegendentry{ACMA};

\end{axis}
\end{tikzpicture}%
}
	\caption{Average SINR of batch BSS algorithms vs. $N_s$ for $N_t=5, N_r=7$, SNR $=30$dB and $N_{Sweeps}=8$.}
	\label{fig:fig6}
\end{figure}
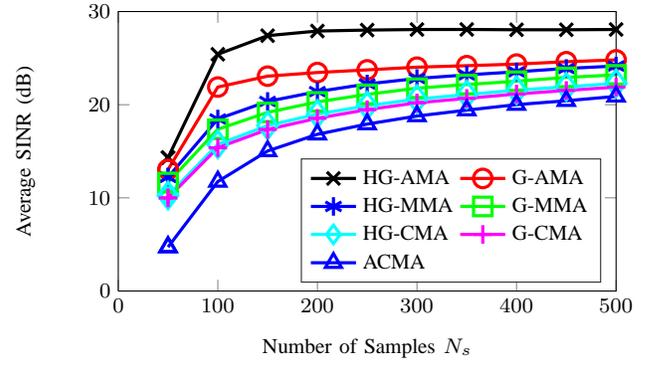
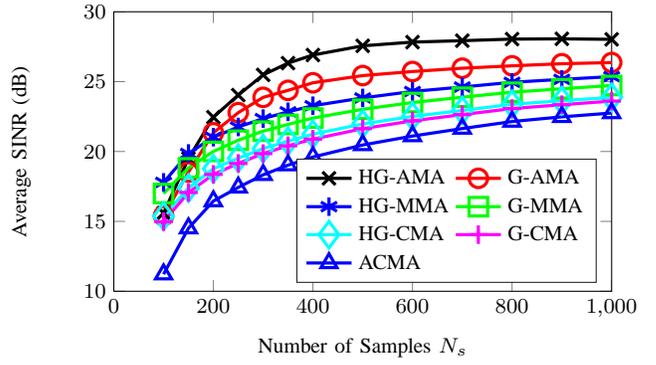

Figure \ref{fig:fig6a} and \ref{fig:fig6b}, show the SINR of our proposed algorithms vs. the number of samples $N_s$ for $64$-QAM and $256$-QAM constellations, respectively. The SNR and the total number of sweeps $N_{Sweeps}$ are fixed at 30 dB, and 8, respectively. We notice that as expected, the larger the number of samples the better the performance is. However, we observe a threshold point after which the gain is not significant as the SINR will be essentially limited by the SNR value. It can be seen that the performance of AM algorithms is better than MM and CM algorithms. Also, HG-AMA takes the lead among all other algorithms.

\subsection{Experiment 7: Comparison based on SER}\label{Exp7}
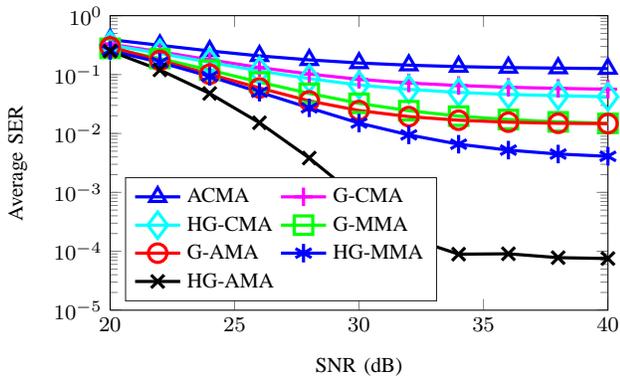
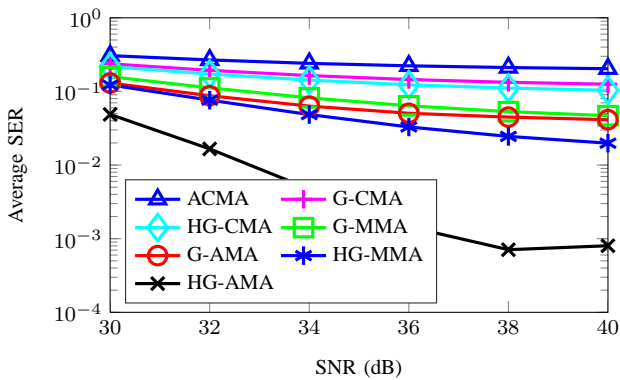
\begin{figure}[tb!]\centering
	\setlength\figureheight{5.5cm}
	\setlength\figurewidth{8.2cm}
    \subfloat[$64$-QAM and $N_s=300$]{\label{fig:fig7a}
	\footnotesize
\definecolor{mycolor1}{rgb}{0.00000,1.00000,1.00000}%
\definecolor{mycolor2}{rgb}{1.00000,0.00000,1.00000}%
\begin{tikzpicture}
\begin{axis}[
width=\figurewidth,
height=\figureheight,
xmin=20,xmax=40,
xlabel={SNR (dB)},
ymode=log,
ymin=1e-05,ymax=1,
yminorticks=true,
ylabel={Average SER},
legend columns=2,
legend style={at={(0.03,0.03)},anchor=south west,legend cell align=left,align=left,draw=white!15!black}
]

\addplot [color=blue,solid,line width=1.2pt,mark size=3.5pt,mark=triangle,mark options={solid}]
  table[row sep=crcr]{%
20	0.392792666666667\\
22	0.314327333333334\\
24	0.252285333333334\\
26	0.2076\\
28	0.177739333333333\\
30	0.157393333333333\\
32	0.144412666666667\\
34	0.136512666666667\\
36	0.131565333333333\\
38	0.128672666666667\\
40	0.126313333333333\\
};
\addlegendentry{ACMA};

\addplot [color=mycolor2,solid,line width=1.2pt,mark size=3.5pt,mark=+,mark options={solid}]
  table[row sep=crcr]{%
20	0.330061333333333\\
22	0.244086666666667\\
24	0.178172\\
26	0.132029333333333\\
28	0.102226666666667\\
30	0.08304\\
32	0.0719100000000001\\
34	0.0652253333333333\\
36	0.0609246666666667\\
38	0.0582973333333333\\
40	0.0565146666666667\\
};
\addlegendentry{G-CMA};

\addplot [color=mycolor1,solid,line width=1.2pt,mark size=5.0pt,mark=diamond,mark options={solid}]
  table[row sep=crcr]{%
20	0.31729\\
22	0.229018\\
24	0.161744\\
26	0.115192\\
28	0.0854693333333334\\
30	0.0664506666666667\\
32	0.056094\\
34	0.0498093333333333\\
36	0.0460246666666666\\
38	0.043836\\
40	0.041926\\
};
\addlegendentry{HG-CMA};

\addplot [color=green,solid,line width=1.2pt,mark size=3.5pt,mark=square,mark options={solid}]
  table[row sep=crcr]{%
20	0.279060666666667\\
22	0.187448666666667\\
24	0.119314\\
26	0.0745846666666666\\
28	0.0479793333333333\\
30	0.03253\\
32	0.0243686666666667\\
34	0.0197206666666666\\
36	0.017236\\
38	0.0157466666666666\\
40	0.0148673333333333\\
};
\addlegendentry{G-MMA};

\addplot [color=red,solid,line width=1.2pt,mark size=3.5pt,mark=o,mark options={solid}]
  table[row sep=crcr]{%
20	0.295076\\
22	0.174386666666667\\
24	0.101508\\
26	0.059244\\
28	0.0357386666666666\\
30	0.024528\\
32	0.0193346666666666\\
34	0.016802\\
36	0.0155606666666667\\
38	0.0148953333333333\\
40	0.0146993333333333\\
};
\addlegendentry{G-AMA};

\addplot [color=blue,solid,line width=1.2pt,mark size=3.5pt,mark=asterisk,mark options={solid}]
  table[row sep=crcr]{%
20	0.255708666666667\\
22	0.161296\\
24	0.0931439999999999\\
26	0.0502986666666667\\
28	0.02724\\
30	0.0150226666666666\\
32	0.00946266666666664\\
34	0.00657799999999998\\
36	0.00522199999999999\\
38	0.00447666666666666\\
40	0.00409666666666667\\
};
\addlegendentry{HG-MMA};

\addplot [color=black,solid,line width=1.2pt,mark size=3.5pt,mark=x,mark options={solid}]
  table[row sep=crcr]{%
20	0.250650666666667\\
22	0.119932666666667\\
24	0.0478046666666666\\
26	0.0152113333333333\\
28	0.00383\\
30	0.000787333333333333\\
32	0.00017\\
34	8.93333333333333e-05\\
36	9.06666666666667e-05\\
38	7.8e-05\\
40	7.53333333333333e-05\\
};
\addlegendentry{HG-AMA};

\end{axis}
\end{tikzpicture}%
} \vfill \medskip
    \subfloat[$256$-QAM and $N_s=900$]{\label{fig:fig7b}
	\footnotesize
\definecolor{mycolor1}{rgb}{0.00000,1.00000,1.00000}%
\definecolor{mycolor2}{rgb}{1.00000,0.00000,1.00000}%
\begin{tikzpicture}
\begin{axis}[
width=\figurewidth,
height=\figureheight,
xmin=30,xmax=40,
xlabel={SNR (dB)},
ymode=log,
ymin=0.0001,ymax=1,
yminorticks=true,
ylabel={Average SER},
legend columns=2,
legend style={at={(0.03,0.03)},anchor=south west,legend cell align=left,align=left,draw=white!15!black}
]

\addplot [color=blue,solid,line width=1.2pt,mark size=3.5pt,mark=triangle,mark options={solid}]
  table[row sep=crcr]{%
30	0.306460222222222\\
32	0.267768\\
34	0.240592666666667\\
36	0.222810888888889\\
38	0.211099111111111\\
40	0.203516888888889\\
};
\addlegendentry{ACMA};

\addplot [color=mycolor2,solid,line width=1.2pt,mark size=3.5pt,mark=+,mark options={solid}]
  table[row sep=crcr]{%
30	0.237526888888889\\
32	0.193480222222222\\
34	0.164257333333333\\
36	0.145128888888889\\
38	0.133003777777778\\
40	0.125355555555556\\
};
\addlegendentry{G-CMA};

\addplot [color=mycolor1,solid,line width=1.2pt,mark size=5.0pt,mark=diamond,mark options={solid}]
  table[row sep=crcr]{%
30	0.217280666666667\\
32	0.172446222222222\\
34	0.142679111111111\\
36	0.12311\\
38	0.111073777777778\\
40	0.103802\\
};
\addlegendentry{HG-CMA};

\addplot [color=green,solid,line width=1.2pt,mark size=3.5pt,mark=square,mark options={solid}]
  table[row sep=crcr]{%
30	0.158735111111111\\
32	0.111984444444444\\
34	0.0819844444444444\\
36	0.0638857777777778\\
38	0.0531995555555555\\
40	0.0469024444444444\\
};
\addlegendentry{G-MMA};

\addplot [color=red,solid,line width=1.2pt,mark size=3.5pt,mark=o,mark options={solid}]
  table[row sep=crcr]{%
30	0.130311555555556\\
32	0.0874228888888889\\
34	0.0632237777777777\\
36	0.0505468888888889\\
38	0.0448302222222222\\
40	0.0413244444444445\\
};
\addlegendentry{G-AMA};

\addplot [color=blue,solid,line width=1.2pt,mark size=3.5pt,mark=asterisk,mark options={solid}]
  table[row sep=crcr]{%
30	0.122489111111111\\
32	0.0762106666666667\\
34	0.0485075555555556\\
36	0.032932\\
38	0.0245046666666667\\
40	0.0198353333333333\\
};
\addlegendentry{HG-MMA};

\addplot [color=black,solid,line width=1.2pt,mark size=3.5pt,mark=x,mark options={solid}]
  table[row sep=crcr]{%
30	0.0487693333333333\\
32	0.0165737777777778\\
34	0.00474111111111111\\
36	0.00145044444444445\\
38	0.000708888888888889\\
40	0.000802222222222222\\
};
\addlegendentry{HG-AMA};

\end{axis}
\end{tikzpicture}%
}
	\caption{Average SER of AM, MM and CM algorithms vs. SNR for $N_t=5$, $N_r=7$, $N_{Sweeps}=8$ and different $N_s$.}
	\label{fig:fig7}
\end{figure}

Figure \ref{fig:fig7a} and \ref{fig:fig7b} depict the SER of AM, MM and CM algorithms vs. SNR for the case of $64$-QAM and $256$-QAM constellations, respectively. The number of samples $N_s=300$ and $N_s=900$ are considered for the case of $64$-QAM and $256$-QAM, respectively. As noticed previously, the performance of the HG-AMA is significantly better than all the other algorithms. Similar to other figures, same pattern of performance is observed i.e., the HG-AMA takes the lead followed by the HG-MMA, G-AMA, G-MMA and then by the HG-CMA, G-CMA and ACMA. By observing these figures, we can say that HG-AMA is the only algorithm which works well for high-order QAM constellations.

\section{Conclusion}\label{sec:conclusion}
In this paper, fundamental problems with the physical layer for MIMO systems are addressed. The targeted problems include channel estimation and blind demixing. Mainly, the problem focussed here is to design algorithms for high-order QAM signals without using pilot symbols. Four new iterative batch BSS algorithms are presented; two of them dealing with the MM criterion namely G-MMA and HG-MMA and the other two dealing with the AM criterion namely G-AMA and HG-AMA. The proposed algorithms are designed using a pre-whitening operation to reduce the complexity of optimization problem, followed by a recursive separation method of unitary Givens and J-unitary hyperbolic rotations for the minimization of MM/AM criteria. Instead of using complex matrices, a real transformation is considered where a special structure of the separation matrix in the whitened domain is suggested and maintained throughout all transformations.

The proposed algorithms are mainly designed for the blind demixing of MIMO systems involving high-order QAM signals. Simulation results demonstrate their favorable performance as compared to the state of the art algorithms dealing with the CM criterion such as G-CMA, HG-CMA and ACMA. It is noticed that the G-MMA and G-AMA are cheaper and more suitable for large number of samples but in the case of small number of samples the HG-MMA and HG-AMA should be used. Moreover, out of all the currently available batch BSS algorithms and the presented ones, the alphabet matched algorithm designed by combining Givens and hyperbolic rotations (HG-AMA) is the most efficient one for high-order QAM signals such as 64-QAM and 256-QAM.

\bibliographystyle{IEEEtran}
\bibliography{MMA_AMA}
\end{document}